\documentclass[a4paper,11pt]{article}
\usepackage{cite}
\usepackage{jheppub} 
\usepackage{etoolbox}
\patchcmd{\maketitle}{\@fpheader}{}{}{}
\usepackage[T1]{fontenc} 
\usepackage[utf8]{inputenc} 
\usepackage{microtype} 
\usepackage{lmodern} 
\usepackage{mdframed} 
\usepackage{mathtools}
\usepackage{graphicx}
\usepackage{cancel} 
\usepackage{todonotes} 
\usepackage[normalem]{ulem} 
\usepackage{soul} 
\usepackage{amsmath}
\usepackage{amsfonts}
\usepackage{amssymb}
\usepackage{stmaryrd}
\usepackage[mathscr]{euscript}
\usepackage{mathtools}
\usepackage{mathrsfs}
\usepackage{multicol}
\usepackage{mleftright} \mleftright
\usepackage{tensor} 
\usepackage{braket} 
\usepackage{mathrsfs} 
\usepackage{bbold}
\usepackage{color}
\usepackage{braket}
\usepackage{slashed}
\usepackage{hyperref}

\newcommand{\tJ}{\tilde J}
\newcommand{\tP}{\tilde P}

\newcommand{\tY}{\tilde Y}

\newcommand{\tLambda}{\tilde \Lambda}
\newcommand{\tmu}{\tilde \mu}

\newcommand{\comment}[1]{}

\providecommand{\sc}{{ }}
\renewcommand{\sc}{{ }}

\DeclareMathAlphabet{\mathfs}{U}{rsfs}{m}{n}                     %

\newcommand{\asl}{$\mathfrak{sl}(2,\mathbb R)$ }
\newcommand{\aslu}{$\mathfrak{sl}(2,\mathbb R)\times\mathbb R $ }

\newcommand{\n}{\nonumber}
\newcommand{\be}{\nopagebreak[3]\begin{equation}}
\newcommand{\ee}{\end{equation}}
\newcommand{\bee}{\nopagebreak[3]\begin{equation*}}
\newcommand{\eee}{\end{equation*}}
\newcommand{\ba}{\nopagebreak[3]\begin{eqnarray}}
\newcommand{\ea}{\end{eqnarray}}
\newcommand{\baa}{\nopagebreak[3]\begin{eqnarray*}}
\newcommand{\eaa}{\end{eqnarray*}}
\newcommand{\bal}{\nopagebreak[3]\begin{aligned}}
\newcommand{\eal}{\end{aligned}}

\newcommand{\bseq}{\nopagebreak[3]\begin{subequations}}
\newcommand{\eseq}{\end{subequations}\noindent}

\title{Non-relativistic and Carrollian limits of Jackiw-Teitelboim gravity}
\author[1]{Joaquim Gomis}
\author[2,3,4]{Diego Hidalgo}
\author[5]{Patricio Salgado-Rebolledo}

\affiliation[1]{Departament de F\'isica Qu\`{a}ntica i Astrof\'isica and Institut de Ci\`{e}ncies del Cosmos (ICCUB), Universitat de Barcelona, Mart\'i i Franqu\`{e}s 1, E-08028 Barcelona, Spain}
\affiliation[2]{Centro de Estudios Cient\'ificos (CECs), Av. Arturo Prat 514, Valdivia, Chile,}
\affiliation[3]{Departamento de F\'isica, Universidad de Concepci\'on, Casilla 160-C, Concepci\'on, Chile,}
\affiliation[4]{Instituto de Ciencias F\'isicas y Matem\'aticas, Universidad Austral de Chile, Edificio Emilio Pugin, cuarto piso, Campus Isla Teja, Valdivia, Chile}
\affiliation[5]{Universit\'e Libre de Bruxelles and International Solvay Institutes, ULB-Campus Plaine CP231, B-1050 Brussels, Belgium}
\emailAdd{joaquim.gomis@ub.edu}
\emailAdd{dihidalgot@gmail.com}
\emailAdd{psalgado@ulb.ac.be}
\preprint{ICCUB-20-024}

\abstract{
We construct the non-relativistic and Carrollian versions of Jackiw-Teitelboim gravity. In the second order formulation,
there are no divergences in the non-relativistic and Carrollian limits. Instead, in the first order formalism there are divergences that can be
avoided by starting from a relativistic BF theory with (A)dS$_2\times \mathbb R$ gauge algebra. We show how to define the boundary duals of the gravity actions using the method of non-linear realisations and suitable Inverse Higgs constraints. In particular, the non-relativistic version of the Schwarzian action is constructed in this way. We derive the asymptotic symmetries of the theory, as well as the corresponding conserved charges and Newton-Cartan geometric structure. Finally, we show how the same construction applies to the Carrollian case. 
}

\begin{document}
\maketitle
\flushbottom

\section{Introduction}
The Sachdev-Ye-Kitaev (SYK) model \cite{Sachdev:1992fk, kitaev2015simple,Kitaev:2017awl} is a solvable quantum mechanical model of Majorana fermions in one dimension.
In the regime of large coupling (low temperature) this model is perturbative in the $1/N$-expansion,
and shows emergent  conformal symmetry, which is the 
reparametrisation symmetry group Diff($S^1$).
The one dimensional diffeomorphism symmetry is
spontaneously broken to $SL(2, R)$, whose low energy dynamics 
is described by the 1D Schwarzian theory \cite{Kitaev:2017awl,Maldacena:2016hyu,Stanford:2017thb}.
The holographic description of SYK model
is the  Jackiw-Teitelboim (JT) gravity \cite{Teitelboim:1983ux,Jackiw:1984je} 
 in two dimensions and the boundary description of the
 bulk JT gravity is also given the Schwarzian action
  \cite{Maldacena:2016upp,Engelsoy:2016xyb,Jensen:2016pah}. The SYK model has been generalised to include complex fermions, its low energy dynamics \cite{Sachdev:2015efa,Davison:2016ngz,Bulycheva:2017uqj,Chaturvedi:2018uov,Gu:2019jub,Godet:2020xpk} 
 is described  by a generalised Schwarzian theory with symmetry 
 $SL(2, R)\times U(1)$.
  A flat limit of this generalised Schwarzian action was studied in 
 \cite{Afshar:2019axx}.
 JT gravity and its boundary description are also useful for
 the momentum/complexity
correspondence  and its connection with the non-relativistic Newton's law 
\cite{Brown:2018bms,Susskind:2019ddc,Barbon:2020olv}.

Motivated by the previous ideas and with the goal of finding another physical sector of the SYK model and its 
holographic description,\footnote{In the case of the original AdS/CFT duality two interesting sectors has been considered the BMN 
\cite{Berenstein:2002jq}
and the one associated to the string non-relativistic limit of 
 $AdS_5\times S^5$ \cite{Gomis:2005pg}. 
}
we will consider the non-relativistic (NR) and Carrollian limit of JT gravity and its boundary action. The starting point of this study consists in writing the JT gravity action as a BF theory with (A)dS$_2$ gauge group \cite{Fukuyama:1985gg,Isler:1989hq,Chamseddine:1989yz}. Then, as shown in \cite{Galajinsky:2019lak}, the Schwarzian action can also be obtained via the method non-linear
  realisations \cite{Coleman}
     and the inverse Higgs mechanism (IHM) \cite{Ivanov:1975zq}. However, in order to have a finite NR limit we will need extend the analysis and consider a BF theory
with gauge group (A)dS$_2\times \mathbb{R}$. The NR and Carrollian bulk actions can also be expressed as BF theories with Newton-Hooke\footnote{
The plus sign corresponds to the contraction of dS algebra while the minus sign corresponds to the contraction of AdS algebra. Note that, the space-time symmetries of the harmonic oscillator form the NH$_2^{+}$ algebra, whereas the inverted harmonic oscillator has a NH$_2^{-}$ symmetry algebra .}  (NH$_2^\pm$) and Carroll (A)dS$_2$ gauge algebras, respectively \cite{Bacry:1968zf,dubouis,Levy-Leblond}. 

To study the NR boundary theory we should consider the conformal basis of (A)dS$_2\times \mathbb{R}$, and introduce two different contractions of the $SL(2,\mathbb R) \times\mathbb R$ algebra. The first one leads to  
\be
[\mathcal H,\mathcal  D] = \mathcal H \,, \qquad [\mathcal K,\mathcal D] = -\mathcal K \,, \qquad [\mathcal H,\mathcal K] = 2\mathcal Z \,,
\ee
which we name the {\it Extended Galilean conformal algebra in 1D dimension}. This algebra is isomorphic to NH$^+$. The second contraction gives
\be
[\hat{\mathcal H},\hat{ \mathcal  D}] = \hat{ \mathcal K} \,, \qquad [\hat{ \mathcal K},\hat{ \mathcal D}] = -\hat{ \mathcal H} \,, \qquad [\hat{ \mathcal H},\hat{ \mathcal K}] = 2\hat{ \mathcal Z} \,,
\ee
which is isomorphic to NH$^-$ and we will refer to it as {\it Twisted Extended Galilean conformal algebra in 1D dimension}. These two algebras are isomorphic as complex algebras.  
 
The boundary actions are constructed via the non-linear realisation method and IHM. In the case of a NR limit of the dS$_2\times \mathbb R$, the boundary action 
 has Extended Galilean conformal symmetry. We name this action {\it Non-Relativistic Schwarzian}. Instead, if we consider the NR limit of the
AdS$_2\times \mathbb R$ algebra, the boundary action has Twisted Extended Galilean conformal symmetry. It is a complex version of the NR Schwarzian which is closely related to the flat Schwarzian action of \cite{Afshar:2019axx}.

The Carrollian limit of the relativistic bulk and boundary actions is obtained from the observation that the Carroll (A)dS$_2$ algebra admits a central extension in the commutator of  the Galilean boost and momentum 
generator\footnote{The existence 
of this extension is a unique feature of the two-dimensional case since, unlike the Galilean case, 
the Carroll algebra does not admit a non-trivial 
central extension in four dimensions.},  
we name  this algebra {\it Extended Carroll (A)dS$_2$ algebra}.
One can see that this symmetry follows from the Extended NH$_2^{\pm}$ algebra by interchanging the generators $H$ and $P$ and changing the sign of the cosmological constant.
This fact allows one to pass from Galilean to Carrollian symmetries\footnote{This relation generalises the known duality between Galilean and Carrollian symmetries \cite{Barducci:2018wuj} in the flat case, see also \cite{Duval:2014uoa,Bergshoeff:2020xhv}.}
\be\label{dualities}
\bal
&{\rm Extended\;\,NH}_2^+\leftrightarrow{\rm Extended\;\, Carroll\; AdS}_2\\
&{\rm Extended\;\, NH}_2^- \leftrightarrow {\rm Extended \;\,Carroll \;dS}_2\,.
\eal
\ee
The Carrollian actions in the bulk and in the boundary can therefore be obtained from the NR ones. 

The IHM allows us to construct the boundary BF gauge fields. Similarly as done in 3D gravity \cite{Banados:1994tn}, it is possible to further reconstruct bulk gauge fields, from which we can construct Newton-Cartan \cite{Cartan:1923zea,Trautman:1963,Havas:1964zza,Kuenzle:1972zw,Kuchar:1980tw}
 and Carrollian geometric structures \cite{Henneaux:1979vn,Duval:2014uoa}. 

The NR and Carrollian limits of the JT bulk action \cite{Teitelboim:1983ux,Jackiw:1984je} 
in the second order formalism are also studied. In this case the divergences in the NR and Carrollian limits can be absorbed by rescaling the Newton constant. This is due to the fact that in two dimensions there are no divergent terms in the expansion of the Ricci scalar, which is a remarkably property that is in high contrast with its analogue in three and four space-time dimensions  \cite{DePietri:1994je,VandenBleeken:2017rij,Hansen:2020pqs,Bergshoeff:2019ctr}.

The organisation of the paper is as follows: in Section \ref{secJT}, we review the main aspects of JT gravity in first and second order formulations. In Section \ref{secNRJT}, we study the NR limit of the JT action, first in the second order formulation, and subsequently in the first order formulation by considering a BF theory with (A)dS$_2\times \mathbb R$ gauge algebra. The Section \ref{secCJT} is devoted to the Carrollian limit of the JT gravity theory, which is carried out by using the duality
among NR and Carrollian symmetries.
Section \ref{schsection} deals with the construction of the boundary theory of JT gravity. Here we provide a derivation of the known Schwarzian theory and its asymptotic symmetries by means of the non-linear realisation method and
IHM. We generalize the procedure to the (A)dS$_2\times \mathbb R$ case. In Section \ref{NRschsection}, we define the NR Schwarzian action and its asymptotic symmetries and conserved charges. We show how this is generalized to the Carrollian case. Finally, in Section \ref{outlook}, we give our conclusions and we elaborate on relevant future directions and possible generalizations of our results.\\

{\bf Note added}:  When this paper was finished, we noticed in the arXiv the article \cite{Grumiller:2020elf}, {\it ``Limits of JT gravity''} by D.~Grumiller, J.~Hartong, S.~Prohazka and J.~Salzer, [arXiv:2011.13870 [hep-th]]. Some of the results of this paper coincide with ours. However, the techniques used in the derivations are different.

\vskip1truecm

\section{Jackiw-Teitelboim gravity}\label{secJT}
In this section we review certain aspects of the Jackiw-Teitelboim (JT) gravity \cite{Teitelboim:1983ux,Jackiw:1984je} which is dual to  the SYK model \cite{Sachdev:1992fk,kitaev2015simple,Kitaev:2017awl} in $(0+1)$-dimensions. We start considering the second order formulation of JT gravity and subsequently review its first order formulation.

\subsection{Jackiw--Teitelboim action}

The JT action for gravity in $(1+1)$-dimensions is given by
\be \label{JT}
S_{\rm JT} \, = \, \tilde \kappa \int  d^2 x\sqrt{-g}\, \Phi (\mathcal R- 2\tLambda) \,,
\ee
where $\mathcal R$ is the Ricci scalar, $\tLambda$ is the cosmological constant, $\Phi$ a scalar field, and $\tilde \kappa$ a two-dimensional coupling constant. The field equations that follow from varying $\Phi$ and $g_{\mu \nu }$ in \eqref{JT} read
\bseq
\ba\label{eqRJT}
\mathcal R  - 2\tLambda & = & 0\,,\\ \label{eqscalarJT}
(\nabla_\mu \nabla_\nu -g_{\mu \nu} \nabla^2) \Phi+\tLambda g_{\mu \nu} \Phi  &=&  0\,,
\ea
\eseq
where $\nabla_\mu$ denotes the affine space-time covariant derivative. The second equation can be decomposed into traceless and trace parts  as \cite{Jackiw:1993gf}
\bseq
\ba
\left(  \nabla_\mu \nabla_\nu -\frac{1}{2} g_{\mu \nu} \nabla^2   \right) \Phi &=  0 \,,  \\
(\nabla^2 -2 \tLambda   )\Phi  &= 0\,, \label{KGAdS}
\ea
\eseq
where we recognize \eqref{KGAdS} as the Klein-Gordon equation on (A)dS$_2$. Using this equation, one can see that \eqref{eqscalarJT} is equivalent to
\be\label{eqscalarJT2}
\nabla_\mu \nabla_\nu \Phi-\tLambda\, g_{\mu \nu} \Phi  =  0\,.
\ee
In the following, we will review how the above geometric dynamics may be presented in a gauge theoretical fashion.

\subsection{First order formulation of JT gravity}

A first order formulation of JT gravity can be defined by gauging the (A)dS$_2$ symmetry and constructing a BF theory \cite{Fukuyama:1985gg,Isler:1989hq,Chamseddine:1989yz}. The gauge algebra is given by $\mathfrak{so}(1,2)$ in the AdS case and $\mathfrak{so}(2,1)$ for dS. The generators satisfy the following commutation relations\footnote{For the conventions see Appendix \ref{AppAdS}.}
\be \label{adS2algebra}
\left[ \tJ, \tP_a \right] =  {\epsilon_a}^b \tP_b \,,  \hskip.9truecm  \left[ \tP_a, \tP_b \right]  =-\tLambda \epsilon_{ab} \tJ\,,
\ee
where $\tP_a$ stand for translations and $\tJ$ is the boost generator. The invariant tensor in (A)dS$_2$ is given by 
\be\label{ads2invt}
\langle \tJ, \tJ \rangle =\tmu  , \hskip.8truecm  \langle \tP_a, \tP_b \rangle =- \tmu\, \tLambda \, \eta_{ab}\,, 
\ee
where $\tmu$ is an arbitrary constant. 

In order to define a BF theory we consider a scalar field $B$ taking values in the (A)dS$_2$ algebra
\be\label{JTB}
 B=\Phi^a\, \tP_a + \Phi\, \tJ\,,
\ee
and a set of one-form gauge fields,  $E^a=E^a_\mu\, dx^\mu$ and 
 $\Omega=\Omega_\mu\, dx^\mu$, corresponding to the zweibein form and the dual spin connection $\Omega\equiv- \frac{1}{2}\varepsilon_{ab} \, \Omega^{ab}$, respectively. The gauge fields define the (A)dS$_2$ connection one form
\be\label{JTconnection}
A=E^a\, \tP_a +\Omega\, \tJ\,,
\ee
together with the corresponding curvature two-form $F=dA +A^2$, given by\footnote{Wedge product between differential forms is assumed.}
\be
F=R^a(\tP)\tP_a+R(\tJ)\tJ\,,\hskip.7truecm
R^a(\tP)=dE^a - \epsilon^a_{\;\;b} \, \Omega E^b \,,\hskip.5truecm
R(\tJ)=d\Omega  -\frac{\tLambda}{2} \epsilon_{ab}E^a E^b \,.
\ee
The BF action then reads
\be \label{JTBFaction}
S[B,A]=\int \left\langle B, F \right\rangle   = \tilde\mu\int \left[\Phi R(\tJ)-\tLambda \Phi_a R^a(\tP) \right]\,,
\ee
and its field equations are given by
\bseq
\ba
&\delta \Phi^a:\hskip.5truecm &R^a(\tP)=0\,,\label{eqT}\\[5pt]
&\delta \Phi:\hskip.5truecm &R(\tJ)=0\,,\label{eqR}\\[5pt]
&\delta E^a:\hskip.5truecm &d\Phi^a-\epsilon^a_{\;\;b}\left(\Phi E^b+\Omega \Phi^b\right)=0\,,\label{eqBa}\\[5pt]
&\delta \Omega:\hskip.5truecm &d\Phi-\tLambda\epsilon_{ab}E^a \Phi^b=0\label{eqB}\,.
\ea
\eseq
From equation \eqref{eqT} we can solve the spin connection in terms of the zweibein as
\be\label{OmegaintermsofE}
\Omega_\mu(E)=-E^{-1}\epsilon^{\alpha\beta}\partial_\alpha E^a_\beta E_{a\mu}\,, 
\ee
where
\be
E\equiv{\rm det}(E^a_\mu)=-\frac{1}{2}\epsilon^{\mu\nu}\epsilon_{ab}E^a_\mu E^b_\nu\,.
\ee
Using this result, one can evaluate the Riemann tensor
\be\label{Riemann}
\mathcal R(E)^{a}_{\;\;b\rho\sigma}=E^a_\mu E^\nu_b \,\mathcal R(E)^{\mu}_{\;\;\nu\rho\sigma}=-\epsilon^{a}_{\;\;b}\bigg(\partial_\rho \Omega(E)_\sigma -
\partial_\sigma \Omega(E)_\rho \bigg)\,,
\ee
which indicates that the field equation \eqref{eqR} can be rewritten in terms of the Ricci scalar $\mathcal R=g^{\mu\nu}\mathcal R^{\rho}_{\;\;\mu\rho\nu}$ as
\be\label{eqRJ}
\epsilon^{\mu\nu}\left(\partial_\mu\Omega(E)_\nu-\frac{\tLambda}{2}\epsilon_{ab}E^a_\mu E^b_\nu \right)=-\frac{E}{2}\left(\mathcal R(E)-2\tLambda\right)=0\,,
\ee
and reproduces the field equation \eqref{eqRJT} found in the second order formulation. On the other hand, from \eqref{eqB} we can express $\Phi^a$ in terms of $\Phi$ as
\be
E^\mu_a \Phi^a= \frac{E^{-1}}{\tLambda}\epsilon^{\mu\nu}\partial_\nu \Phi\,.
\ee
Using this equation in \eqref{eqB} yields
\be\label{eqPhirel}
\nabla_\mu \nabla_\nu \Phi-\tLambda g_{\mu \nu} \Phi  =  0\,,
\ee
where the space-time metric is defined in the usual way
 \be \label{spMetric}
 g_{\mu\nu} \, = \, E^a_\mu\, E^b_\nu \, \eta_{ab}\,.
 \ee
This reproduces the field equation \eqref{eqscalarJT2} previously found in the second order formulation.
Using \eqref{eqT} and \eqref{eqRJ}, the BF action takes the form of the JT action \eqref{JT}, namely
\be\label{reducedJT}
S[B,A]\, = \,-\frac{\tmu}{2}\int d^2x E\,\Phi(\mathcal R(E)-2\tLambda)\,,
\ee
when $\tmu$ and $\tilde \kappa$ are properly identified.

However, since the spin connection was not solved by means of its own field equation, the reduced action \eqref{reducedJT} is not dynamically equivalent to the original BF action \eqref{JTBFaction}. Thus, the equivalence
between the first order and second order formulations of JT gravity should be considered only at the level of their field equations. The solutions in the second-order formalism are contained as subset of the solutions of the first-order formalism.

\section{Non-relativistic Jackiw-Teitelboim gravity}\label{secNRJT}
In this section, we will consider the NR limit of JT gravity in second and first order formulations. In the first order formulation, the limit is applied to a BF theory with gauge group (A)dS$_2\times\mathbb R$. 

\subsection{Second order formulation}
The NR expansion of JT gravity follows in general terms from the approaches studied in \cite{Dautcourt:1997hb,Weinbergbook, dautcourt1964newtonske,DePietri:1994je,Hansen:2020pqs,VandenBleeken:2017rij,Ehlers:2019aco,Kuenzle:1972zw}. Let us consider the power expansion of the relativistic metric $g_{\mu \nu} (x)$ and $g^{\mu \nu} (x)$ up to $\varepsilon^2$-order, as follows
\be \label{transgmunu}
g_{\mu \nu} =  \frac{1}{\varepsilon^2} \, {\overset{(2)} g}_{\mu \nu} +  {\overset{(0)} g}_{\mu \nu} + \mathcal{O}\left( \varepsilon^2  \right)\,, \hskip.8truecm
g^{\mu \nu} =  {\overset{(0)} g}{}^{\mu \nu}+ \varepsilon^2 \, {\overset{(-2)} g}{}^{\mu \nu} + \mathcal{O}\left( \varepsilon^4 \right)\,, \\
\ee
where the higher inverse powers of $\varepsilon=1/c \rightarrow 0$ correspond to Post-Newtonian corrections that will not be considered here. From the relation $g_{\mu\rho}\, g^{\rho\nu}=\delta^\nu_\mu$, we obtain the following relations for the different terms in the expansion
\be \label{conditionsgmunu}
{\overset{(2)} g}_{\mu \nu} 
{\overset{(-2)} g}{}^{\nu \rho}
+{\overset{(0)} g}_{\mu \nu} 
{\overset{(0)} g}{}^{\nu \rho}
= \delta^{\rho}_{\mu}, \qquad 
{\overset{(2)} g}_{\mu \nu} 
{\overset{(0)} g}{}^{\nu \rho}=0, 
\qquad 
{\overset{(0)} g}_{\mu \nu} 
{\overset{(-2)} g}{}^{\nu \rho}
=0\,.
\ee 
Now we introduce a symmetric affine Levi-Civita connection with the expansion\footnote{Notice that here we could consider a more general affine connection, including an antisymmetric part related to the torsion in Lorentzian geometry, but we do not consider that case.} 
\be\label{Gammaexpansion}
\Gamma^{\lambda}_{\mu \nu}  =  \frac{1}{\varepsilon^2} \,  \overset{(2)}{\Gamma}{}^{\lambda}_{\mu \nu} + \overset{(0)}{\Gamma}{}^{\lambda}_{\mu \nu} + \varepsilon^2 \, \overset{(-2)}{\Gamma}{}^{\lambda}_{\mu \nu}\,.
\ee
As usual, we demand metric compatibility on the metric and its inverse
\ba\label{metricompa}
\nabla_\rho \, g_{\mu \nu}  \, =  \, 0 \,, \qquad\qquad \nabla_\rho \, g^{\mu \nu}  \, =  \,  0\,,
\ea
where the covariant derivative $\nabla_\rho$ is defined with respect to the symmetric connection \eqref{Gammaexpansion}. For the covariant metric $g_{\mu \nu}$, this condition implies the following set of equations order by order
\bseq
\ba 
\overset{(0)}{\nabla}_\rho    \overset{(0)}{g}_{\mu \nu}  -\overset{(-2)}{\Gamma}{}^{\lambda}_{\rho \nu} \overset{(2)}{g}_{\mu \lambda} -\overset{(-2)}{\Gamma}{}^{\lambda}_{\rho \mu} \overset{(2)}{g}_{\nu \lambda} & = & 0\,, \\
 \overset{(0)}{\nabla}_\rho     \overset{(2)}{g}_{\mu \nu} -\overset{(2)}{\Gamma}{}^{\lambda}_{\rho \nu} \overset{(0)}{g}_{\mu \lambda} -\overset{(2)}{\Gamma}{}^{\lambda}_{\rho \mu} \overset{(0)}{g}_{\nu \lambda} & = & 0 \,, \\
 \overset{(2)}{\Gamma}{}^{\lambda}_{\rho \nu} \overset{(2)}{g}_{\mu \lambda} +\overset{(2)}{\Gamma}{}^{\lambda}_{\rho \mu} \overset{(2)}{g}_{\nu \lambda}  & = & 0 \,, \\
 \overset{(-2)}{\Gamma}{}^{\lambda}_{\rho \nu}  \overset{(0)}{g}_{\mu \lambda}+ \overset{(-2)}{\Gamma}{}^{\lambda}_{\rho \mu}  \overset{(0)}{g}_{\nu \lambda}  & = & 0\,,
\ea 
\eseq
where we have introduced the covariant derivative $\overset{(0)}{\nabla}:= \partial + \overset{(0)}{\Gamma}$. Solving the system yields (see Appendix \ref{NRGammasapp})
\bseq\label{connections}
\ba
 &\overset{(2)}{\Gamma}{}^{\lambda}_{\mu \nu} & = \frac{1}{2} \,  {\overset{(0)}g}{}^{\lambda \sigma}  \left(  \partial_\nu {\overset{(2)}g}_{\sigma \mu} + \partial_\mu {\overset{(2)}g}_{\sigma \nu} - \partial_\sigma {\overset{(2)}g}_{\mu \nu}   \right)
\,, \label{Gamma2}\\[5pt]
\n &\overset{(0)}{\Gamma}{}^{\lambda}_{\mu \nu} & =   \frac{1}{2}{\overset{(-2)}g}{}^{\lambda \sigma}  \left(  \partial_\nu {\overset{(2)}g}_{\sigma \mu} + \partial_\mu {\overset{(2)}g}_{\sigma \nu} - \partial_\sigma {\overset{(2)}g}_{\mu \nu}   \right)
\\&&\;+\, \frac{1}{2}\,{\overset{(0)}g}{}^{\lambda \sigma}  \,\left(  \partial_\nu {\overset{(0)}g}_{\sigma \mu} + \partial_\mu {\overset{(0)}g}_{\sigma \nu} - \partial_\sigma {\overset{(0)}g}_{\mu \nu}   \right)
\,, \\[5pt]
 &\overset{(-2)}{\Gamma}{}^{\lambda}_{\mu \nu} & =  \frac{1}{2}  \, {\overset{(-2)}g}{}^{\lambda \sigma}  \left(  \partial_\nu {\overset{(0)}g}_{\sigma \mu} + \partial_\mu {\overset{(0)}g}_{\sigma \nu} - \partial_\sigma {\overset{(0)}g}_{\mu \nu}   \right)\,.
\ea
\eseq
Now we write the metric \eqref{spMetric} in terms of Newton-Cartan fields $\tau_\mu$ and $ e_\mu$, and the vector field $m_\mu$ by considering the zweibein expansion \cite{Cartan:1923zea,Trautman:1963,Havas:1964zza,Kuenzle:1972zw, Kuchar:1980tw}
\be\label{veilbein2ndorder}
E^{0}_\mu =  \frac{1}{\varepsilon} \, \tau_\mu + \frac{\varepsilon}{2} \, m_\mu \,,\hskip.8truecm
E^{1}_\mu =   e_\mu\,,
\ee
where $m_\mu$ is the vector field associated to the central extension of the Bargmann algebra \cite{DePietri:1994je,Andringa:2010it}. This leads to
\be\label{tmunuhmunu}
{\overset{(2)}g}_{\mu \nu}  = -  \tau_\mu  \, \tau_\nu \,,\hskip.8truecm
{\overset{(0)}g}_{\mu \nu} =   h_{\mu\nu} - \tau_{(\mu} \, m_{\nu )}\,,
\ee
where we have introduced the spatial metric $h_{\mu\nu}\equiv \,e_\mu e_\nu$. The relations $E^\mu_a \, E_\nu^a =\delta^\mu_\nu $, and $E^\mu_b \, E_\mu^a=\delta^a_b$ lead to the following inverse zweibeine
\be\label{inspiredcontraction}
E_{0}^\mu = \varepsilon  \tau^\mu  -\frac{\varepsilon^3}{2} \, \tau^\mu \tau^\nu m_\nu \,,\hskip.8truecm
E_{1}^\mu =   e^\mu-\frac{\varepsilon^2}{2}\tau^\mu e^\nu m_\nu\,,
\ee
which involve the new additional fields $\tau^\mu$ and $e^\mu$ satisfying
the relations
\be\label{S1invertconditions}
e^\mu e_\nu +\tau^\mu \tau_\nu=\delta^\mu_\nu, \qquad \tau^\mu  e_{\mu }= 0, \quad   e^{\mu }\tau_\mu= 0, \quad  \tau_\mu  \tau^\mu =1, \quad e^\mu e_\mu=1\,.
\ee
Note that the spatial metric $h_{\mu \nu}$ is degenerate. With these definitions the inverse metric takes the form 
\be\label{inverseg}   
\bal
g^{\mu \nu}  &= \eta^{ab} E^\mu_{a}  E^\nu_{b} \,,\\ 
&= e^\mu e^\nu -\varepsilon^2\tau^\mu\tau^\nu  -\frac{\varepsilon^2}{2} \left( \tau^\mu e^\nu +e^\mu \tau^\nu  \right)e^\sigma m_\sigma
+\varepsilon^4 \tau^\mu \tau^\nu e^\sigma e^\rho m_\sigma m_\rho\,.
\eal
\ee 
Therefore, considering terms only up to order $\varepsilon^2$, we obtain the following form of the inverse temporal and spatial metrics in \eqref{transgmunu}, 
\be\label{tmunuhmunuinv}
{\overset{(0)}g}{}^{\mu \nu}  =e^\mu  e^\nu\equiv  h^{\mu\nu}\,, \hskip.8truecm
{\overset{(-2)}g}{}^{\mu \nu} =-\tau^{\mu}\tau^{ \nu} -    \tau^{(\nu } h^{\mu)\sigma} m_\sigma  \,,
\ee
where we have introduced the inverse spatial metric $h^{\mu\nu}$. With these explicit forms for the metric components, the terms in the connection expansion \eqref{connections} take the form
\be\label{explicitGammas}
\begin{aligned}
\overset{(2)}{\Gamma}{}^{\lambda}_{\mu \nu}&=   h^{\lambda \sigma} \left( \tau_\mu \partial_{[\sigma} \tau_{\nu]} + \tau_\nu \partial_{[\sigma} \tau_{\mu]}  \right) \,, \\
\overset{(0)}{\Gamma}{}^{\lambda}_{\mu \nu} &=   \tau^\lambda  \partial_{(\mu} \tau_{\nu)}  + \frac{1}{2} h^{\lambda \sigma} \left( \partial_\nu h_{\sigma \mu} + \partial_\mu h_{\sigma \nu} - \partial_\sigma h_{\mu \nu} \right)+h^{\lambda\sigma}\left(\partial_{[\sigma} m_{\mu]} \tau_{\nu}+\partial_{[\sigma} m_{\nu]} \tau_{\mu}\right)\,, \\
\overset{(-2)}{\Gamma}{}^{\lambda}_{\mu \nu} & =  - \frac{1}{2} [ \tau^\lambda \tau^\sigma +\tau^{(\lambda } h^{\sigma)\rho} m_\rho  ] \left(  \partial_{\nu} [ e_\sigma e_\mu -\tau_{(\sigma} m_{\mu)}   ]   \right. \\
&\hskip3.6cm \left. + \partial_{\mu} [ e_\sigma e_\nu - \tau_{(\sigma} m_{\nu)}   ] - \partial_{\sigma} [ e_\mu e_\nu - \tau_{(\mu} m_{\nu)}   ]      \right) \,.
\end{aligned}
\ee
Given \eqref{transgmunu}, the corresponding expansion of the Ricci scalar $\mathcal{R}$ is (see Appendix \ref{DivergentRicciscalar})
\be\label{Ricciexpansion}
\mathcal{R} \, = \, \varepsilon^2  \, {\overset{\rm(-2)}{ \mathcal R }} + \mathcal{O} \left( \varepsilon^4  \right) \,.
\ee
We see that in two dimensions there are no divergent terms in the expansion of the Ricci scalar. This is a remarkably property that is in high contrast with three and four space-time dimensions  \cite{DePietri:1994je,VandenBleeken:2017rij,Hansen:2020pqs,Bergshoeff:2019ctr}.

We consider now the expansion of the vielbein postulate,
\be \label{relativisticpostulate}
 \partial_\mu {E_\lambda}^a + {\varepsilon^a}_{b}\, \Omega_\mu \, E_\lambda^b - {\Gamma}^{\rho}_{\mu \lambda} E_\rho^{a}  \, = \, 0\,.
\ee
Implementing \eqref{Gammaexpansion}, \eqref{inspiredcontraction} and considering
\be\label{omegaexp}
\Omega_\mu =\displaystyle   \varepsilon \, \omega_\mu\,,
\ee
the anti-symmetric part of \eqref{relativisticpostulate} is
\begin{multicols}{2}
\begin{subequations}\label{symeqgamma}
\setlength{\abovedisplayskip}{-13pt}
\allowdisplaybreaks
\begin{align}
\partial_{[\mu}  \, \tau_{\lambda]} & =  0\,,\\[.2truecm]
 \partial_{[\mu} e_{\lambda]} +  \omega_{[\mu}  \tau_{\lambda]} &  = 0\,,\\[.1truecm]\label{eqforomega11}
\partial_{[\mu} m_{\lambda ]} + \omega_{[\mu} e_{\lambda]}  & =   0\,,\\[.1truecm]\label{eqforomega2}
 \omega_{[\mu}  m_{\lambda]} & =  0 \,,
\end{align}
\end{subequations}
\end{multicols}\noindent
whereas the symmetric part reads
\bseq\label{antisymeqgamma}
\ba
\overset{(2)}{\Gamma}{}^\rho_{\mu \lambda} \, \tau_\rho &  =&0 \,,\\
\overset{(2)}{\Gamma}{}^\rho_{\mu \lambda} \, e_\rho&  =&0 \,,\\
\partial_{\mu} \tau_{\lambda } -\overset{(2)}{\Gamma}{}^\rho_{\mu \lambda} m _\rho-\overset{(0)}{\Gamma}{}^\rho_{\mu \lambda} \, \tau_\rho  & = & 0\,,\\
 \partial_{\mu} m_{\lambda } + \omega_{\mu} e_{\lambda}-\overset{(0)}{\Gamma}{}^\rho_{\mu \lambda} m_\rho  - \overset{(-2)}{\Gamma}{}^\rho_{\mu \lambda} \,\tau_\rho &  = & 0\,, \\
 \partial_{\mu} e_{\lambda} +  \omega_{\mu} \tau_{\lambda}- \overset{(0)}{\Gamma}{}^\rho_{\mu \lambda} e_\rho &= &  0\,, \\
 \omega_{\mu} m_{\lambda}-\overset{(-2)}{\Gamma}{}^\rho_{\mu \lambda}\,e_\rho  & = & 0\,. 
\ea 
\eseq
Solving this system for all the 
 connections, yields 
\ba\label{gammasfromveilbeingpostulate}
 \overset{(2)}{\Gamma}{}^\rho_{\mu \lambda}  & = &  0\,, \\
 \overset{(0)}{\Gamma}{}^\rho_{\mu \lambda}  & = & \tau^\rho \, \partial_{\mu} \tau_{\lambda } +e^\rho\, \left(\partial_{\mu} e_{\lambda } +\omega_{\mu} \tau_{\lambda} \right)\,, \\
 \overset{(-2)}{\Gamma}{}^\rho_{\mu \lambda}  & = & \tau^\rho \left[\partial_{\mu} m_{\lambda } + \omega_{\mu} e_{\lambda}-\left( \tau^\sigma \, \partial_{\mu} \tau_{\lambda } +e^\sigma\, \left(\partial_{\mu} e_{\lambda } +\omega_{\mu} \tau_{\lambda} \right)\right) m_\sigma  \right] + e^\rho \omega_{\mu} m_{\lambda} \,.
\ea
which is compatible with \eqref{explicitGammas}. Using this result, the leading term in the Ricci scalar expansion \eqref{Ricciexpansion} takes the form (see Appendix \ref{AppendixSeconOrder} for details),
\be\label{R-2NR}
 \mathcal {\overset{\rm(-2)}R} = 4\, \tau^\mu e^\nu \,\partial_{[\mu}\, \omega_{\nu ]}\,,
\ee
where the NR spin connection $\omega_\mu$ is solved algebraically by equations \eqref{eqforomega11} and \eqref{eqforomega2}, yielding
\be\label{spinconnection2order}
 \omega_\mu =  2\, \tau^{[\alpha}e^{\beta ]} \,\left(  e_\mu \partial_\alpha e_\beta- \tau_\mu \partial_\alpha m_\beta  \right).
\ee
Finally, we expand the metric determinant, the JT scalar field, the cosmological constant, and the coupling constant as 
\bseq\label{determinantS1}
\ba
\sqrt{-g} =\det \left( E^a_\mu \right) & = &  -\epsilon^{\mu \nu} {E_\mu}^0 {E_\nu}^1  =\frac{1}{\varepsilon} \det (\tau e) + \mathcal{O} \left( \varepsilon\right)\,,\\ \label{rescalings}
\Phi & = &  \varepsilon  \, \phi \,,\hskip.8truecm \tilde\Lambda=\varepsilon^2 \, \Lambda \hskip.8truecm \tilde\kappa={\frac{1}{\varepsilon^2}} \, \kappa\,.
\ea
\eseq
Note that the rescaling of $\tilde\kappa$ corresponds to a rescaling of the Newton's constant $G$. Then, in the $\varepsilon\rightarrow0$ limit, we find the NR action,
\be\label{JTNR}
S_{\scriptscriptstyle NRJT}= \kappa \int d^2 x \, \det \left(\tau e \right) \, \phi\, \left(\mathcal {\overset{\rm(-2)}R}-2\Lambda\right),
\quad \det (\tau e) \equiv -\epsilon^{\mu \nu} \tau_\mu e_\nu\,.
\ee
This action leads to the second order equation
\ba
\mathcal R^{\scriptscriptstyle NR} -2\Lambda=0\,, 
\ea
where we have defined the NR Ricci scalar curvature
\be
\mathcal R^{\scriptscriptstyle NR} \equiv \mathcal {\overset{\rm(-2)}R}=
8\tau^{[\mu} e^{\nu]}\partial_{\mu}\left(\tau^{[\alpha} e^{\beta]}\left(\partial_{\alpha}e_{\beta}e_{\nu}-\partial_{\alpha}m_{\beta}\tau_{\nu} \right)\right)\,.
\ee
This defines the NR version of the relativistic JT equation \eqref{eqRJT}. Unlike second order gravitational actions in higher dimensions, \eqref{JTNR} has no divergent terms. As we will see in the next section, the rescalings \eqref{rescalings} are compatible with the first-order formulation results. We will also show that the field equations of the second-order formalism are a submanifold of the field equations of the first order formalism.

\subsection{First order formulation}\label{FirstOrderFormulationSec}

The NR limit of JT gravity in the first-order formulation can be defined starting from a BF theory with gauge algebra (A)dS$_2\times  \mathbb R $. Thus we extend the (A)dS$_2$ symmetry \eqref{adS2algebra} by including an Abelian generator $\tY$ and consider the invariant bilinear form
\be\label{invtenZ}
\langle \tJ, \tJ \rangle =\tmu  , \hskip.8truecm  \langle \tP_a, \tP_b \rangle =- \tmu\, \tLambda \, \eta_{ab}\,, \qquad \langle \tY,\tY\rangle=-\,\tmu   \,\tilde \Lambda\,.
\ee
The NR contraction follows from defining NR generators $H$, $P$, $J$ and $M$ as
\begin{multicols}{2}
\begin{subequations}\label{NRgenerators}
\setlength{\abovedisplayskip}{-13pt}
\allowdisplaybreaks
\begin{align}
\tP_0 & =  \frac{\varepsilon}{2}  H +  \frac{1}{\varepsilon} \, M\,,\\[.2truecm]
 \tP_1 & =  P\,,\\[.1truecm]
\tJ  &=  \frac{1}{\varepsilon} G \,,\\[.1truecm]
 \tY  &=  \frac{\varepsilon}{2}  H -\frac{1}{\varepsilon}\, M\,.
\end{align}
\end{subequations}
\end{multicols}\noindent
 Inverting this relation, we find
\begin{multicols}{2}
\begin{subequations}\label{NRgeneratorsInv}
\setlength{\abovedisplayskip}{-13pt}
\allowdisplaybreaks
\begin{align}
H& = \frac{1}{\varepsilon}\, \left(\tP_0 +\tY \right)\,,\\[.2truecm]
 P& =  \tP_1\,,\\[.1truecm]
G &=   \varepsilon   \tJ \,,\\[.1truecm]
 M  &=\frac{\varepsilon}{2} \left(\tP_0 -\tY \right)\,.
\end{align}
\end{subequations}
\end{multicols}\noindent
Using the commutation relations \eqref{adS2algebra}, and defining the NR cosmological constant $\Lambda$ as
\be\label{rescLambda0}
\Lambda = \frac{1}{\varepsilon^2} \tLambda\,,
\ee
we find, in the the limit $\varepsilon \rightarrow 0$, the centrally Extended Newton-Hooke algebra \cite{Bacry:1968zf,dubouis} in two space-time dimensions:
\ba \label{newton-hookealgebra}
\left[ G , H\right]  = P, \quad \quad  \left[ G,  P\right]  = M , \quad \quad  \left[ H, P\right]  =  -\Lambda\, G.
\ea
Using the invariant tensor on (A)dS$_2$ \eqref{ads2invt} together with \eqref{invtenZ}, and defining
\be\label{rescmu}
\mu= \varepsilon^2 \, \tilde\mu \,,
\ee
the contraction \eqref{NRgenerators} yields the following NR invariant bilinear form
\be \label{NRinvten}
\langle G , G \rangle =   \mu\,,\quad \langle H , M \rangle  =    \mu \, \Lambda\,, \quad \langle P, P \rangle  = -   \mu \,\Lambda\,,
\ee 
which is non-degenerate for $\mu\neq 0$.
We next consider the NR limit 
of the JT action in first-order formulation \eqref{JTBFaction}. With this aim we consider a one-form connection taking values on (A)dS$_2\times \mathbb{R} $ algebra expressed in terms relativistic and NR generators \cite{Bergshoeff:2015uaa}
\be\label{relnonrelconnection}
A= E^a \tP_a +\Omega \tJ + X \tY=
\tau H + e P + \omega G + m M
\ee
where, using \eqref{NRgenerators}, we find
\begin{multicols}{2}
\begin{subequations}\label{firstcontractionfield}
\setlength{\abovedisplayskip}{-13pt}
\allowdisplaybreaks
\begin{align}
E^{0} & =  \frac{1}{\varepsilon} \, \tau + \frac{\varepsilon}{2} \, m \,,\\[.2truecm]\label{E0}
E^{1} & =   e\,,\\[.1truecm]\label{E1}
\Omega & =   \varepsilon \,\omega\,,\\[.1truecm]\label{omegabigJT}
X& =  \frac{1}{\varepsilon} \, \tau - \frac{\varepsilon}{2} \, m \,.
\end{align}
\end{subequations}
\end{multicols}\noindent
Besides, we extend the definition of the scalar field \eqref{JTB} to take values on (A)dS$_2\times \mathbb{R}$ as well
\be\label{JTBext}
 B=\Phi^a\tP_a + \Phi \tJ +  \Psi \tY\,,
\ee
and define the NR scalar fields $\{\eta, \rho,\phi, \zeta\}$
\begin{multicols}{2}
\begin{subequations}\label{NRlagrangefield}
\setlength{\abovedisplayskip}{-13pt}
\allowdisplaybreaks
\begin{align}
\Phi^0 & = \frac{1}{\varepsilon} \eta + \frac{\varepsilon}{2} \zeta \,,\\[.2truecm]
\Phi^1 & =  \rho \,,\\[.1truecm]
\Phi & =   \varepsilon \, \phi\,,\label{phileading} \\[.1truecm]
\Psi & =   \frac{1}{\varepsilon} \eta - \frac{\varepsilon}{2} \zeta\,.
\end{align}
\end{subequations}
\end{multicols}\noindent
Using these definitions, the expansion of the relativistic BF action for the (A)dS$_2 \times \mathbb R$ algebra is
\be \label{NRaction0}
\bal
&S[  B, A] = \tilde\mu\int \left(\Phi R(\tJ)-\tLambda \Phi_a R^a(\tP) \right) -\,\tmu   \,\tilde \Lambda \int \Psi \, dX\\
&= \varepsilon^2 \, \tilde \mu \int \Bigg(\phi R(G) +\Lambda\bigg(\eta R (M) +\zeta R (H)-\rho R (P)\bigg)\Bigg) + \frac{\varepsilon^4 \, \Lambda \tilde\mu}{2 }\int \left(\zeta \omega e -\phi m e -\rho \omega m \right)\,,
\eal
\ee
where we have used \eqref{rescLambda0} and \eqref{invtenZ}, and we have defined the NR curvature two-forms as
\begin{multicols}{1}
\begin{subequations}\label{NRcurvatures}
\setlength{\abovedisplayskip}{-13pt}
\allowdisplaybreaks
\begin{align}
R(H) & =  d\tau     \,,\\[.1truecm]
R(P ) &=  d e +\omega\tau\,,\\[.1truecm]
R(G)  &=  d  \omega -\Lambda \tau e
\,,\\[.1truecm] 
R(M)  &=   d m + \omega e\,.
\end{align}
\end{subequations}
\end{multicols}\noindent
Using \eqref{rescmu} and taking the limit $\varepsilon \rightarrow 0$, we obtain the NR two-dimensional gravity theory
\ba \label{NRaction0lim}
S = \mu\int \Bigg(\phi R(G) +\Lambda\bigg(\eta R (M) +\zeta R (H)-\rho R (P)\bigg)\Bigg) \,.
\ea
It is worth to mention that this action can be alternatively obtained as a BF theory based on the Extended Newton-Hooke algebra
(\ref{newton-hookealgebra}),
 where the gauge connection $A$ and the $B$ field are given by
\be
\bal
A&= \tau H + e P + \omega G + m M\,,\\
B&= \eta H + \rho P + \phi G + \eta M\,.
\eal
\ee
In this case the curvature two form associated to $A$ takes the form
\be
R= R(H) \,H + R(P) \, P+ R(G)\, G +R(M)\, M\,,
\ee
with the components given in \eqref{NRcurvatures}. It is straightforward to see that, using these definitions, the BF action 
 leads exactly to \eqref{NRaction0lim} when using the NR invariant tensor \eqref{NRinvten} 

The field equations coming from the action \eqref{NRaction0lim} when varying with respect to $\eta$, $\rho$, $\phi$ and $\zeta$ are given by
\bseq\label{fieldeqS1R}
\ba
&\delta \eta: \qquad R(H) & =  0\,, \label{fieldeqRHS1.1} \\[5pt]
&\delta \rho: \qquad R(P) &=  0 \,,\label{fieldeqRS1}\\[5pt]
&\delta \phi: \qquad R(G) &= 0\,, \label{fieldeqRGS1}\\[5pt]
&\delta \zeta: \qquad R(M)  &=  0\,, \label{fieldeqRHS1} 
\ea
\eseq
while varying \eqref{NRaction0lim} with respect to $\tau$, $e$, $\omega$ and $m$ leads to
\bseq\label{fieldeqS1sf}
\ba
\delta \tau:& \qquad &d\eta  = 0\,, \label{fieldeqRHS1} \\[4pt]
\delta e:& \qquad &d\rho
+
\omega \eta
- \tau \phi 
 =  0 \,,\label{fieldeqRPS1}\\[4pt]
\delta \omega:& \qquad &d\phi
-\Lambda 
\left(
\tau \rho
-e \eta
\right)
  = 0\,, \label{fieldeqRG}  \\[4pt]
 \delta m: & \qquad &d\zeta +
\omega \rho
- e \phi
  =  0\,. \label{fieldeqRMS1} 
 \ea
\eseq
Note that equation \eqref{fieldeqRHS1.1} implies
\be
d\tau=0 \quad\Longrightarrow \quad\tau=d\lambda\,,
\ee
for some zero-form $\lambda$, therefore on-shell there is no torsion. This fact implies that the Newton-Cartan structure admit absolute time. Solving the field equations \eqref{fieldeqRPS1} and \eqref{fieldeqRMS1} the spin-connection is given by \eqref{spinconnection2order}. Therefore, the action (\ref{NRaction0lim}) becomes
\ba \label{NRaction0lima}
S = \mu\int \,\phi \, R(G) \, = \, - \mu\int \det{(\tau e)} \phi \Bigg( \mathcal R^{\scriptscriptstyle NR}
-2\Lambda \Bigg)\,.
\ea
This action matches with \eqref{JTNR} identifying $\mu=-\kappa$.

Since the spin connection has not been solved by its own field equation, this action is dynamically inequivalent to the first order action \eqref{NRaction0lim}. This is in complete analogy to the relativistic case.



\section{Carrollian Jackiw-Teitelboim  gravity}\label{secCJT}

In this section, we work out the Carrollian limit of the JT gravity. 
We consider the Carroll contraction of the (A)dS$_2\times \mathbb R$ algebra. In order to do that we first interchange in  $P_0\leftrightarrow P_1$, and $H\leftrightarrow P$ in \eqref{NRgenerators}, which gives
\begin{multicols}{2}
\begin{subequations}\label{Carrollgeneratorsexpansion}
\setlength{\abovedisplayskip}{-13pt}
\allowdisplaybreaks
\begin{align}
\tP_0 & =    H \,,\\[.2truecm]
 \tP_1 & = \frac{\varepsilon}{2} P+  \frac{1}{\varepsilon}  M\,,\\[.1truecm]
\tJ  &=   \frac{1}{\varepsilon}  G \,,\\[.1truecm]
 \tY  &=  \frac{\varepsilon}{2}  P - \frac{1}{\varepsilon}  M\,.
\end{align}
\end{subequations}
\end{multicols}\noindent
 Secondly, taking the limit $\varepsilon \rightarrow0$, and using the rescaling \eqref{rescLambda0}, we get
\ba \label{carroladsalgebra}
\left[ G , P\right]  = H, \qquad  \left[ G,  H\right]  = M , \qquad  \left[ H, P\right]  = - \Lambda \,G.
\ea
We shall call this symmetry {\it Extended Carroll (A)dS$_2$ algebra}. 
One can see that these commutation relations follow from the Extended NH$^{\pm}$ algebra \eqref{newton-hookealgebra} by interchanging $H$ and $P$ and changing the sign of the cosmological constant.
This fact allows one to pass from Galilean to Carrollian symmetries. This relation is a generalisation in two dimensions
of a more general duality between of these two types of symmetries \cite{Barducci:2018wuj} previously found in the flat case. 

 This procedure defines the following dualities
\be\label{dualities}
\bal
&{\rm Extended\;\,NH}_2^+\leftrightarrow{\rm Extended\;\, Carroll\; AdS}_2\\
&{\rm Extended\;\, NH}_2^- \leftrightarrow {\rm Extended \;\,Carroll \;dS}_2\,.
\eal
\ee
It is important to remark that in $1+1$ dimensions, the Carroll algebra (even without cosmological constant) admits the central extension $M$ given in \eqref{carroladsalgebra} in the same way as its Galilean counterpart does. This situation is a unique feature of the two-dimensional case since, unlike the Galilean case, the Carroll algebra does not admit a non-trivial central extension in four dimensions.

\subsection{Second order formulation}
We start carrying out the Carrollian contraction in the second order formulation. 
To make contact with the Carrollian geometry we first
write the relativistic zweibein in terms of the Carrollian zweibein gauge fields
\be\label{veilbeincarroll1}
E^{0}_\mu =  \tau_\mu \,,\hskip.9truecm
E^{1}_\mu =  \frac{1}{\varepsilon} e_\mu +\frac{\varepsilon}{2} m_\mu\,.
\ee
The completeness relations $E^\mu_a \, E_\nu^a =\delta^\mu_\nu $, and $E^\mu_b \, E_\mu^a=\delta^a_b$ imply the inverse relativistic zweibein
\be\label{veilbeincaroll2}
E_{0}^\mu =   \tau^\mu-\frac{\varepsilon^2}{2}e^\mu \tau^\nu m_\nu\,,\hskip.8truecm
E_{1}^\mu = \varepsilon\, e^\mu  -\frac{\varepsilon^3}{2} \, e^\mu e^\nu m_\nu \,.
\ee
In analogy with \eqref{Carrollgeneratorsexpansion}, these expressions can be obtained by interchanging $E^0$ and $E^1$, and interchanging $\tau$ and $e$ in \eqref{veilbein2ndorder}. The gauge fields $\tau, e$ and $m$ verifying the completeness relation \eqref{S1invertconditions}.

From the zweibein we can read the metric expansion
 \be \label{Carrolltransgmunu}
g_{\mu \nu} =  \frac{1}{\varepsilon^2} \, {\overset{(2)} g}_{\mu \nu} +  {\overset{(0)} g}_{\mu \nu} \,, 
\ee
where
\be 
{\overset{(2)}g}_{\mu \nu}  =e_\mu  \, e_\nu \,,\hskip.8truecm
{\overset{(0)}g}_{\mu \nu} =   -\tau_\mu \tau_\nu + e_{(\mu} \, m_{\nu )}\,.
\ee
Using the expression of the inverse zweibein \eqref{veilbeincaroll2}, we find the following form of the inverse metric
\be\label{inversegcarroll}
\bal
g^{\mu \nu}  &= -\tau^\mu \tau^\nu+ \varepsilon^2 \left(h^{\mu \nu} + \tau^{(\mu} e^{\nu)} \tau^\sigma m_\sigma\right)  -\varepsilon^4 \left(   h^{\mu \nu} e^\sigma m_\sigma + \frac{1}{4} h^{\mu \nu} \tau^\sigma \tau^\rho m_\sigma m_\rho   \right)\,.
\eal
\ee 
This has the general form
 \be \label{Carrolltransgmunuinv}
g^{\mu \nu} =  {\overset{(0)} g}{}^{\mu \nu}+ \varepsilon^2 \, {\overset{(-2)} g}{}^{\mu \nu} + \mathcal{O}\left( \varepsilon^4 \right)\,, \\
\ee
where the components read
\be\label{tmunuhmunuinv}
{\overset{(0)}g}{}^{\mu \nu}  = -\tau^\mu \tau^\nu  \hskip.8truecm
{\overset{(-2)}g}{}^{\mu \nu} =h^{\mu \nu} + \tau^{(\mu} e^{\nu)} \tau^\sigma m_\sigma \,,
\ee
with the inverse degenerate spatial metric defined as $h^{\mu \nu} =e^\mu e^\nu.$

Note that \eqref{Carrolltransgmunu} and \eqref{Carrolltransgmunuinv} have the same form than the NR expansion given in the expression \eqref{transgmunu}. However, in this case, the expansion parameter is defined as $\varepsilon =c $. Therefore, we will understand as Carrollian contraction the limit $\varepsilon \rightarrow 0$, which corresponds to the ultra-relativistic limit. With this consideration, the relations \eqref{conditionsgmunu}-\eqref{connections} also hold in this case.

The expansion \eqref{Gammaexpansion} of the Carrollian affine connection can be obtained from the NR one \eqref{gammasfromveilbeingpostulate} by interchanging the gauge fields $\tau$ and $e$. In a similar way to the NR second-order formulation, the first non-vanishing term in the expansion of Ricci scalar \eqref{Ricciexpansion} is given by 
\be\label{RicciCarrollian}
 \mathcal R^{\scriptscriptstyle Carrollian}\equiv \mathcal {\overset{\rm(-2)}R} = 4\, e^\mu \tau^\nu \,\partial_{[\mu}\, \omega_{\nu ]}\,,
\ee
where the Carrollian spin connection $\omega_\mu$ is 
\be\label{spinconnCarroll}
 \omega_\mu =  2\, e^{[\alpha}\tau^{\beta ]} \,\left(  \tau_\mu \partial_\alpha \tau_\beta- e_\mu \partial_\alpha m_\beta  \right)\,.
\ee
As before the Carrollian limit for the JT action \eqref{JT}, can be read off from \eqref{determinantS1} by interchanging $\tau$ and $e$. Finally, taking the limit $\varepsilon\rightarrow0$, the Carrollian version of JT gravity is given by
\be\label{CarrollJTaction}
S_{\scriptscriptstyle Carrollian \,JT}= \kappa \int d^2 x \, \det \left(\tau e \right) \, \phi\, \left( \mathcal R^{\scriptscriptstyle Carrollian}-2\Lambda\right)\,.
\ee
Notice that as in the NR case the divergent terms has been cancelled. This is a special property of two-dimensions.

\subsection{First order formulation}
 
 In this section, we consider the Carrollian limit of the JT gravity in the first-order formalism. Let us consider a one-form connection taking values on (A)dS$_2\times \mathbb{R}$ algebra expressed in terms relativistic and Carrollian generators 
\be\label{relnonrelconnection}
A= E^a \tP_a +\Omega \tJ + X \tY=
e \,H + \tau \, P + \omega \,G + m\, M\,,
\ee
where, using \eqref{Carrollgeneratorsexpansion}, we find 
\begin{multicols}{2}
\begin{subequations}\label{firstcontractionfield}
\setlength{\abovedisplayskip}{-13pt}
\allowdisplaybreaks
\begin{align}
E^{0} & =  \tau \,,\\[.2truecm]\label{E0}
E^{1} & =   \frac{1}{\varepsilon}e+\frac{\varepsilon}{2}m  \,,\\[.1truecm]\label{E1}
\Omega & =   \varepsilon \,\omega\,,\\[.1truecm]\label{omegabigJT}
X& =  \frac{1}{\varepsilon} \, e - \frac{\varepsilon}{2} \, m \,.
\end{align}
\end{subequations}
\end{multicols}\noindent
Also we consider the scalar field to take values on (A)dS$_2\times \mathbb{R}$ 
\be\label{JTBextCarroll}
 B=\Phi^a\tP_a + \Phi \tJ +  \Psi \tY\,,
\ee
and define the Carrollian scalar fields $\{\eta, \rho,\phi, \zeta\}$ as
\begin{multicols}{2}
\begin{subequations}\label{NRlagrangefield}
\setlength{\abovedisplayskip}{-13pt}
\allowdisplaybreaks
\begin{align}
\Phi^0 & = \eta  \,,\\[.2truecm]
\Phi^1 & =\frac{1}{\varepsilon} \rho+  \frac{\varepsilon}{2} \zeta  \,,\\[.1truecm]
\Phi & =   \varepsilon \, \phi\,,\label{phileading} \\[.1truecm]
\Psi & = \frac{1}{\varepsilon} \rho-  \frac{\varepsilon}{2} \zeta \,.
\end{align}
\end{subequations}
\end{multicols}\noindent
These expressions can also be obtained from \eqref{NRlagrangefield} by interchanging $\Phi^0 \leftrightarrow \Phi^1$, and $\eta \leftrightarrow \rho$. Using these definitions, the relativistic BF action takes the form
\be \label{Carollaction0}
\bal
&S[  B, A] = \tilde\mu\int \left(\Phi R(\tJ)-\tLambda \Phi_a R^a(\tP) \right) -\,\tmu   \,\tilde \Lambda \int \Psi\, dX\\
&= \varepsilon^2 \, \tilde \mu \int \Bigg(\phi R(G) +\Lambda\bigg(\rho R (M) +\zeta R (P)-\eta R (H)\bigg)\Bigg) + \frac{\varepsilon^4 \, \Lambda \tilde\mu}{2 }\int \left(\zeta \omega \tau -\phi m \tau -\eta \omega m \right)\,,
\eal
\ee
where we have used \eqref{rescLambda0} and \eqref{invtenZ}, and we have defined the Carrollian curvature two-forms as
\begin{multicols}{1}
\begin{subequations}\label{Carrollcurvatures}
\setlength{\abovedisplayskip}{-13pt}
\allowdisplaybreaks
\begin{align}
R(H) & =  d\tau+\omega e     \,,\\[.1truecm]
R(P ) &=  d e \,,\\[.1truecm]
R(G)  &=  d  \omega -\Lambda e \tau
\,,\\[.1truecm] 
R(M)  &=   d m +\omega  \tau\,.
\end{align}
\end{subequations}
\end{multicols}\noindent
Using \eqref{rescmu}, the Carrollian limit is then obtained by taking $\varepsilon \rightarrow 0$, we find
\ba \label{Carollactionfirstorder}
S = \mu\,  \int \Bigg(\phi R(G) +\Lambda\bigg(\rho R (M) +\zeta R (P)-\eta R (H)\bigg)\Bigg) \,.
\ea
This action can be alternatively obtained as a BF theory based on the Extended Carroll (A)dS$_2$ algebra \eqref{carroladsalgebra} and the bilinear form
\be
\langle G, G \rangle \, = \, \mu \,, \qquad  \langle P, M \rangle \, = \, -\mu\, \Lambda \,,\qquad  \langle H, H \rangle \, = \, \mu\, \Lambda\,,
\ee
which is non-degenerate for $\mu \neq 0$. The field equations coming from the action \eqref{Carollactionfirstorder} when varying with respect to $\eta$, $\rho$, $\phi$, and $\zeta$ read
\bseq
\ba
&\delta \eta: \qquad R(H) & =  0\,, \label{fieldeqRPS1caroll} \\
&\delta \rho: \qquad R(M) &=  0 \,,\\\label{fieldeqRomegacaroll}
&\delta \phi: \qquad R(G) &= 0\,, \\
&\delta \zeta: \qquad R(P)  &=  0\,, 
\ea
\eseq
while varying \eqref{Carollactionfirstorder} with respect to $\tau$, $e$, $\omega$ and $m$ leads to
\bseq
\ba
\delta e:& \qquad &d\rho = 0\,,\\[4pt]
\delta \tau :& \qquad &d\eta + \omega \rho - e \phi =  0 \,,\label{fieldeqRPS1}\\[4pt]
\delta \omega:& \qquad &d\phi
+\Lambda 
\left(
\tau \rho
-e \eta
\right)
  = 0\,, \label{fieldeqRG}  \\[4pt]
 \delta m: & \qquad &d\zeta +
\omega \eta
- \tau \phi
  =  0\,. \label{fieldeqRMS1} 
 \ea
\eseq
Note that equation \eqref{fieldeqRPS1caroll} implies
\be
de=0 \quad\Longrightarrow \quad e=d\lambda\,,
\ee
for some zero-form $\lambda$. This defines an absolute one-dimensional space. Solving the field equations \eqref{fieldeqRPS1} and \eqref{fieldeqRMS1} yields \eqref{spinconnCarroll}. The action \eqref{Carollactionfirstorder}  becomes
\ba \label{NRaction0lima}
S \, =\, \mu\int \,\phi \, R(G)\, = \, - \mu\int \det{(\tau e)} \phi \Bigg( \mathcal R^{\scriptscriptstyle Carrollian}
-2\Lambda \Bigg)\,,
\ea
with the Carrollian curvature $R^{\scriptscriptstyle Carrollian}= $ given in \eqref{RicciCarrollian}.

\section{Relativistic boundary theory}
\label{schsection}

In the previous sections, we have considered the NR and Carrollian limits of JT gravity, focusing only on bulk dynamics and without paying attention to boundary terms. However, in order for the theory to have a well-defined variational principle as well as to define the holographic dual theory, we need to study the boundary dynamics. To address this problem, we consider general BF action of the form 
\be\label{BFactiongen}
S[B,A]\, = \,\int_{ \mathcal M} \left\langle B, F \right\rangle \,,
\ee
where $F=dA+A\wedge A$. For now we consider the zero-form $B$ and the one-form $A$ as taking values on a generic gauge Lie algebra $\mathcal{G}$ with a non-degenerate invariant bilinear form. In order to have a well defined variational principle, we follow the Regge-Teitelboim procedure \cite{Regge:1974zd} and supplement the action \eqref{BFactiongen} with a boundary term
\be\label{JTBFactionB}
S[B,A]\,=\,\int_{ \mathcal M} \left\langle B, F \right\rangle  
+\int_{\partial  \mathcal M}  b\,,
\ee
where $b$ is a one-form defined in such a way that, provided suitable boundary conditions for the gauge fields, the variation of the action does not drop boundary terms. A general variation of \eqref{JTBFactionB} leads to
\be\label{genvarS}
\delta S=\int_{ \mathcal M} 
\bigg(
 \left\langle \delta B, F \right\rangle  
 -\left\langle DB, \delta A \right\rangle  
\bigg)
+\int_{\partial \mathcal M} \bigg( \delta b + \left\langle B, \delta A \right\rangle  \bigg)\,,
\ee
where $D=d + [A, \, \cdot \,]$ is the covariant derivative.
Considering the bulk coordinates $x^\mu=(t,r)$ and the boundary defined at $r\rightarrow\infty$, the gauge fields can be decomposed as
\be
A=A_t \,dt + A_r\, dr \,,\hskip1.2truecm b=b_t dt\,.
\ee
Thus, the boundary term on the right hand side of \eqref{genvarS} takes the form
\be
\int du \bigg( \delta b_t + \left\langle B, \delta A_t \right\rangle  \bigg)\,.
\ee
For this term to vanish, the boundary term $b$ must be such that
\be 
\delta b_t= - \left\langle B, \delta A_t \right\rangle \bigg|_{\partial \mathcal M}\,.
\ee
A boundary condition on the field $B$ that allows to integrate the equation is \cite{Saad:2019lba}
\be\label{bcond}
B\bigg|_{\partial \mathcal M} = k\,  A_t \qquad\Longrightarrow\qquad b_t=-\frac{k}{2} \left\langle A_t , A_t \right\rangle \,,
\ee
with $k$ an arbitrary constant.
On the other hand, the field equations coming from \eqref{genvarS} imply that $B$ is covariantly constant, while $A$ is locally pure gauge, i.e. $A=g^{-1}  dg$ with $g$ an element belonging to a gauge group $\mathcal G$. Furthermore, we consider the following gauge fixing condition \cite{Banados:1994tn}
\be
\partial_t A_r=0 \;\Longrightarrow\; g(t,r) \, = \, U(t)\,b(r)\,.
\ee
The gauge connection then takes the form
\be\label{Aanda}
A\, = \,b^{-1}db + b^{-1} a b \,,
\ee
where $a=U^{-1} dU$ only depends on the coordinate $t$. Evaluating the action \eqref{JTBFactionB} on this solution space, the theory is reduced to the boundary and the action reads
\be\label{boundaryaction2}
S_{\rm bdy}[U] \, = \,-\frac{k}{2}\int dt \left\langle U^{-1} U^{\prime} , U^{-1}  U^{\prime} \right\rangle \,,
\ee
where the prime stands for derivative with respect to $t$. An important step in our construction is that at this point we can consider the boundary target coordinate $t$ as a world-line parameter
of a particle, and the connection $a$ as the pull-back $\star$ of the left-invariant Maurer-Cartan (MC) form $\Omega$ on $\mathcal G$
\be\label{aOmega}
a\, = \, \Omega^\star\,,
\ee
with $\Omega$ satisfying the MC equations
\be
d\Omega+\Omega\wedge\Omega \, = \,0\,.
\ee
Therefore, the boundary action \eqref{boundaryaction2} can be interpreted as the action of a particle moving on the group manifold of $\mathcal G$
\be\label{boundaryaction3}
S_{\rm bdy}[U] \, = \,-\frac{k}{2}\int  d t \, \langle \Omega, \Omega\rangle^{\star}  \,.
\ee
Given a specific form of the connection \eqref{aOmega}, one can look for the residual gauge transformations
\be\label{tgpres}
\delta_\lambda a \, = \, d \lambda + [a  ,\lambda ] \,.
\ee
that preserve its form. The corresponding conserved charges can then be computed using the known result for BF theory \cite{Frodden:2019ylc, Grumiller:2013swa}
\be
\delta Q[\lambda_{bulk}] \, =\, -  \int \langle \lambda_{bulk}, \delta B \rangle d \tau_{\scriptscriptstyle E}\,,
\ee
where we have switched to Euclidean time $\tau_{\scriptscriptstyle E}$ in order to have a periodic coordinate. Imposing the boundary condition \eqref{bcond} for the field $B$ and using the usual decomposition $\lambda_{bulk}= b(r)^{-1} \lambda(t) b(r)$, we find the expression
\be\label{deltaQgen}
\delta Q[\lambda] \, =\, - \, k \int \langle \lambda, \delta a_t \rangle d \tau_{\scriptscriptstyle E}\,,
\ee
Provided the boundary charges can be integrated, the charge algebra is obtained from 
the transformation law
\cite{Regge:1974zd,Banados:1994tn}
\be\label{poissonbracket}
 \delta_{\lambda} Q[\bar \lambda] \,= \, \{ Q[\bar \lambda] , Q[\lambda] \}.
\ee
In the following, we will apply this procedure in the relativistic case and subsequently in the different NR and Carrollian scenarios.

\subsection{(A)dS$_2$ case}
As a first example we consider the (A)dS$_2$ symmetry and the derivation of the Schwarzian action \cite{Kitaev:2017awl,Maldacena:2016hyu,Stanford:2017thb}.
The AdS$_2$ and the dS$_2$ cases can be treated in a unified way by going to the conformal basis, as both they are isomorphic to the \asl algebra (see Appendix \ref{AppAdS})
\be  \label{sl2R}
 [\tilde{ \mathcal{H}},\tilde{ \mathcal {D}}] = \tilde{\mathcal{ H}} \,, \qquad [\tilde{\mathcal{ K}}, \tilde{\mathcal{ D}}] = -\tilde{\mathcal{ K}} \,, \qquad [\tilde{\mathcal{H}}, \tilde{\mathcal{ K}}] = 2\, \tilde{\mathcal{ D}}\,.
 \ee
In this basis, the non-degenerate invariant bilinear form reads
\be\label{pairingsl2R}
\langle \tilde{\mathcal{D}}, \tilde{\mathcal{D}}\rangle= \gamma_0\,, \qquad  \langle \tilde{\mathcal{H}}, \tilde{\mathcal{K}}\rangle= -2\gamma_0,
\ee
with $\gamma_0$ an arbitrary constant. Then, it is possible to locally parametrize an arbitrary group element $\tilde U$ as
 \be\label{groupelementschwarzian}
  \tilde{U} \, = \,  e^{\tilde \rho \tilde{\mathcal{ H}}} \,  e^{\tilde y \tilde{\mathcal{ K}}} \, e^{\tilde u \tilde{\mathcal{D}}}  \,,
 \ee
with $ \tilde \rho, \tilde y ,$  and $\tilde u$ the group manifold coordinates. They will be the Goldstone fields when we take the pull-back to the world-line of the particle. To simplify the computation of the MC one-form $\tilde \Omega$, we use the following $2\times 2$ representation of the $\mathfrak{sl}(2,\mathbb{R})$ algebra
\ba\label{sl2Rirrep} 
 \tilde{\mathcal {H}}  \,=\,  \begin{pmatrix}
0 & 0\\
1& 0
\end{pmatrix}  \,, \qquad  \tilde{\mathcal{ D}} =\frac{1}{2} \begin{pmatrix}
1 & 0\\
0& -1
\end{pmatrix}   \,, \qquad
 \tilde{\mathcal{ K}}  =  \begin{pmatrix}
0 & -1\\
0& 0
\end{pmatrix}    \,.
\ea 
Locally we can parametrize the group element in \eqref{groupelementschwarzian} as
\be
\tilde{U} \, =\,  \left(
\begin{array}{cc}
 \displaystyle e^{\frac{\tilde{u}}{2}} &\quad  -\tilde{y} \, e^{-\frac{\tilde{u}}{2}} \\[1em]
 \tilde{\rho }\, e^{\frac{\tilde{u}}{2}} &\quad  e^{-\frac{\tilde{u}}{2}} \left(1-\tilde{\rho } \,\tilde{y}\right) \\
\end{array}
\right)\,.
\ee
The MC one-form $\tilde \Omega= \tilde{U}^{-1} d \tilde{U}$ is given in a matrix representation by
\be\label{MCformsl2R}
\tilde \Omega \, = \,  \tilde{ \Omega}_{\tilde{\mathcal {H}}  }\, \tilde{\mathcal {H}} + \tilde \Omega_{\tilde{\mathcal {D}}  } \,\tilde{\mathcal {D}} + \tilde \Omega_{\tilde{\mathcal {K}}  }\, \tilde{\mathcal {K}} \, = \, \begin{pmatrix}
\frac{1}{2}  { \tilde{\Omega}}_{ {\mathcal {\tilde{D}}}  } &-  { \tilde{\Omega}}_{ {\mathcal {\tilde{K}}}  } \\[1em]
 { \tilde{\Omega}}_{ {\mathcal {\tilde{H}}}  } & -\frac{1}{2}  { \tilde{\Omega}}_{ {\mathcal {\tilde{D}}}  } 
 \end{pmatrix}\,,
\ee
where we have defined the one-forms, 
\be
\tilde \Omega_{\tilde{\mathcal{H}}} \, =\, e^{\tilde u} \, d \tilde \rho \,,\qquad 
 \tilde \Omega_{\tilde{\mathcal{K}}}  \,=\, e^{-\tilde{u}} \left(\tilde{y}^2\, d\tilde{\rho} +d\tilde{y}\right)   \,,\qquad
\tilde \Omega_{\tilde{\mathcal{D}}} \,=\, d \tilde u +2\tilde y \, d \tilde \rho  \,.
\ee
The MC one-forms can be used to construct the boundary action \eqref{boundaryaction3},\be\label{SL2Raction}
S[U] = -\frac{k}{2}\int dt \,  \langle \tilde \Omega,\tilde \Omega\rangle^{\star}  = -\frac{k\gamma_0}{2} \, \int dt \,\left(4 \tilde \rho' \left(\tilde y \,\tilde u'-\tilde y'\right)+\tilde u'^2\right)\,,
\ee
where prime denotes derivative with respect to the world-line parameter $t$. This model is invariant under the global symmetries
\be\label{globalsymmSch1}
\delta \tilde \rho \,=\, \tilde \alpha +\tilde \beta  \tilde \rho -\tilde \gamma  \tilde \rho ^2 \,, \qquad \delta \tilde y\,=\,(2 \tilde \gamma \tilde \rho-\tilde \beta) \tilde y -\tilde \gamma \,, \qquad \delta \tilde u\,=\, 2 \tilde \gamma  \tilde \rho-\tilde \beta\,,
\ee
where the symmetry parameters can be  grouped as
\be
 \tilde \epsilon \, =\, \tilde \alpha \, \tilde{\mathcal{H}}+ \tilde \beta \, \tilde{\mathcal{D}}+ \tilde \gamma \, \tilde{\mathcal{K}} \, = \, \begin{pmatrix}
\frac{1}{2}   \tilde{\beta}  &-   \tilde{\gamma} \\[1em]
  \tilde{\alpha}& -\frac{1}{2}   \tilde{\beta}
 \end{pmatrix}\,.
\ee
We identify in \eqref{globalsymmSch1} the transformation for the Goldstone field $\tilde \rho$ as the infinitesimal version of a $SL(2,R)$ M\"obius transformation.

We can reduce the number of Goldstone fields in the action by setting some MC forms to zero or defining (invariant)covariant relations among them. This procedure is also know as the {\it inverse Higgs mechanism} (IHM) \cite{Ivanov:1975zq}. In the following, we present an example of this method.\footnote{ We do not attempt to classify all possible IHM constraints.} We consider the constraints
\be\label{constraintSL2R}
\tilde \Omega_{\tilde{\mathcal{H}}} \,=\, \mu \,, \qquad \tilde \Omega_{\tilde{\mathcal{D}}} \,=\,0\,, 
\ee
with $\mu$ a constant. This allows us to express $\tilde y$ and $\tilde u$ in terms of $\tilde \rho$ as follows,
\be\label{constraintSL2R}
 \tilde u=\log \left(\frac{\mu }{\tilde{\rho}^{\prime}}\right) \,, \qquad \tilde y= -\frac{\tilde{u}^{\prime}}{2 \tilde{\rho}^{\prime}}\,,
\ee
for some constant $\mu.$ Let us evaluate  
the MC form on the constraints \eqref{constraintSL2R}
\be\label{reducedMC}
a\,\equiv\,\tilde \Omega^\star \bigg\rvert_{\scriptscriptstyle IHM}  \, = \, \mu \,\tilde{\mathcal {H}}   + L(\tilde \rho) \, \tilde{\mathcal {K}}  \, = \, \begin{pmatrix}
0& \quad -L(\tilde \rho )\\
\mu & \quad 0
 \end{pmatrix}\,,
\ee
where we have introduced the function $L(\tilde \rho) := \frac{1}{2 \, \mu} $Sch$(\tilde \rho, t)$, with 
\be
\text{Sch}(\tilde \rho,t) \, = \, \frac{\tilde{\rho}'''}{ \tilde \rho'} -\frac{3}{2} \left( \frac{\tilde \rho''}{\tilde \rho'} \right)^2\,,
\ee
which is the Schwarzian derivative.
In this case, the IHM is equivalent to a Drinfeld-Sokolov reduction \cite{Drinfeld:1984qv,Polyakov:1989dm}.
Plugging \eqref{constraintSL2R} into the action \eqref{SL2Raction}, we find 
\be\label{OneSchw}
S[\tilde \rho] \, = \, k\gamma_0 \int dt\, \text{Sch} (\tilde \rho,t)\,, 
\ee
which corresponds to the Schwarzian mechanical model invariant under a global $SL(2,R)$ transformation
\be
\tilde \rho_\epsilon \,= \, \frac{a \, \tilde \rho +b}{ c\, \tilde \rho +d} \,, \qquad a\, b -c\, d \, = \,1\,, \qquad \Longrightarrow  \qquad S[\tilde \rho_\epsilon] \, = \, S[\tilde \rho]\,.
\ee
 This result shows that it is possible to obtain the Schwarzian action using the IHM.\footnote{The  Schwarzian can also be obtained by integrating out the gauge transformations of
 the particle model with variables $x^\mu, \lambda$ and Lagrangian $L=\frac 12 \dot x^2-\frac 12 \lambda x^2$, see \cite{Siegel:1988ru,Gomis:1993pp}.}
 
In order to obtain the gauge field in the bulk, we can introduce a radial dependence by performing a gauge transformation with the group element
\be
b\,=\,\begin{pmatrix}
e^{r/2} &0 \\
0 &e^{-r/2}  
\end{pmatrix}\,,
\ee
this leads to the following form for the connection \eqref{Aanda}
\be
A\, = \,b^{-1}db + b^{-1} \Omega^\star b 
\,= \,\begin{pmatrix}
\frac{dr}{2}+\frac{1}{2}  { \tilde{\Omega}^\star}_{ {\mathcal {\tilde{D}}}  } & -e^{-r} \, { \tilde{\Omega}^\star}_{ {\mathcal {\tilde{K}}}  } \\[1.5em]
e^{r} \, { \tilde{\Omega}^\star}_{ {\mathcal {\tilde{H}}}  } &-\frac{dr}{2} -\frac{1}{2}  { \tilde{\Omega}^\star}_{ {\mathcal {\tilde{D}}}  } \end{pmatrix}   \,,
\ee
where we have used \eqref{aOmega} that implements the pull-back of the MC forms to the boundary coordinate $t$. Choosing $\mu=\frac{1}{2}$, we find the the gauge field given in \cite{Saad:2019lba}, namely
\be\label{Amurelativistic}
A=
\begin{pmatrix}
\frac{dr}{2}& 0 \\[1em]
0&-\frac{dr}{2}
\end{pmatrix}
+\frac{dt}{2}
\begin{pmatrix}
0& \qquad -2\, {\rm Sch}( \tilde \rho, t)\,e^{-r}\\[1em]
e^{r}   &\qquad 0
\end{pmatrix}
\,.
\ee
This shows that IHM relations can be interpreted as boundary conditions on the BF connection $A$. Using the change of variables that relates
 \asl and the (A)dS$_2$ algebra (see Appendix \ref{AppAdS}), we can write down the explicit form for the gravitational gauge fields in \eqref{JTconnection}. In the AdS$_2$ case we find
\be
\bal
E^0&=\left(\frac{1}{2}e^r- {\rm Sch}(  \tilde \rho, t)e^{-r} \right)dt\,,\\
E^1&=dr\,,\\
\Omega& =\left(\frac{1}{2}e^r +{\rm Sch}(  \tilde \rho, t) e^{-r}\right)dt \,.\\
\eal
\ee
In order to write down the zweibein and the spin connection for the dS$_2$ case, we interchange the coordinates $t$ and $r$. This is due to the fact that the de-Sitter space has a boundary at time-like infinity. In this particular case, the prime denotes derivative with respect to $r$. We find
\be\label{esomegads}
\bal
E^0 &=dt\,,\\
E^1&=\left(\frac{1}{2}e^t-{\rm Sch}(  \tilde \rho, r) e^{-t}\right)dr\,,\\
\Omega& =\left(\frac{1}{2}e^t +{\rm Sch}(  \tilde \rho, r)e^{-t}\right)dr \,.\\
\eal
\ee
From the expression of the zweibeine, we find the following the space-time metrics
\ba
ds^2_{\scriptstyle AdS} &= &  - \left( \frac{1}{4} e^{2r} -  {\rm Sch}(  \tilde \rho, t) +e^{-2r} \, {\rm Sch}(  \tilde \rho, t)^2  \right) dt^2+dr^2\,, \label{AdSmetric-Case1Schwarzian}\\
ds^2_{\scriptstyle dS} &= &-dt^2 +  \left( \frac{1}{4} e^{2t} -  {\rm Sch}(  \tilde \rho, r) +e^{-2t} \, {\rm Sch}(  \tilde \rho, r)^2  \right) dr^2\,. \label{dSmetric-Case1Schwarzian}
\ea
If we consider the leading term of \eqref{AdSmetric-Case1Schwarzian} for $r \rightarrow \infty$ we recover the boundary conditions of \cite{Saad:2019lba}. 

The asymptotic symmetry is the residual symmetry that leaves the conditions \eqref{reducedMC} invariant. These conditions restrict the gauge parameter 
\be\label{gaugeparameterSL2R}
\lambda  \, =\, \lambda_0  \, \tilde{\mathcal{H}} +  \lambda_1  \, \tilde{\mathcal{K}} +  \lambda_2  \, \tilde{\mathcal{D}}\, =\, \begin{pmatrix}
\frac{1}{2}   \lambda_2  &-   \lambda_1 \\[1em]
  \lambda_0 & -\frac{1}{2}   \lambda_2
 \end{pmatrix}\,.
\ee
  The resulting gauge parameter is given by
\be
\lambda = \lambda_0 \, \tilde{\mathcal{H}}+ \left(\mathcal L \, \lambda_0+\frac{1}{2} \lambda_{0}^{\prime \prime}  \right) \tilde{\mathcal{K}} -\lambda_{0}^{\prime} \, \tilde{\mathcal{D}}\,,
\ee 
whereas the transformation law of the function $L$ leads to
\be\label{variationL}
\delta \mathcal L = \lambda_0 \, \mathcal L^{\prime} + 2\mathcal L \, \lambda_{0}^{\prime} + \frac{1}{2} \, \lambda_{0}^{\prime \prime \prime }\,,
\ee
which is the transformation of a stress tensor under infinitesimal conformal transformations. Using \eqref{deltaQgen}, the associated charge reads
\be\label{integratedcharge}
  Q[\lambda_0] \, = \, 2\gamma_0 \, k \int \lambda_0 \,  \mathcal L  \,d \tau_{\scriptscriptstyle E}\,.
\ee
Then, the Poisson bracket \eqref{poissonbracket}, leads to the Virasoro algebra\footnote{Here prime denotes derivative with respect to the periodic coordinate $\tau_{\scriptscriptstyle E} $.} 
\be\label{Poisson1}
\{  Q[\lambda_0],Q[\bar{\lambda}_0] \} \,  =\, Q[[\lambda_0,\bar{\lambda}_0] ]+\gamma_0 k \int \lambda_0 \bar{\lambda}_{0}^{\prime \prime \prime } d \tau_{\scriptscriptstyle E}\,,
\ee
where $[x,y]=xy'-yx'$ stands for the Lie bracket, and we have changed to Euclidean time in \eqref{variationL}. Introducing the Fourier modes
\be
\mathcal L_m \equiv Q[e^{im \tau_{\scriptscriptstyle E}}] \,, \qquad m \in \mathbb Z\,,
\ee
the Poisson algebra \eqref{Poisson1} takes the usual form 
\be\label{Virasoro}
 \{ \mathcal L_m , \mathcal L_n \} \, = \, (m-n) \mathcal L_{m+n} +\frac{c_1}{ 12 } \,m^3 \, \delta_{m+n,0}\,, \qquad c_1 = 24 \pi \gamma_0 k\,,
\ee
where we have considered the following representation for the Kronecker delta $\delta_{mn} \, =\, \frac{1}{2\pi}  \int e^{i(m-n)\tau_{\scriptscriptstyle E} } d \tau_{\scriptscriptstyle E}$.

\subsection{(A)dS$_2\times \mathbb R$ case}
We turn now our attention to the (A)dS$_2\times \mathbb R$ algebra where the gauge connection is given by 
\be\label{relnonrelconnection0}
A\, =\, E^a \,\tP_a +\Omega\, \tJ + X\, \tY \,,
\ee
In the conformal basis (see Appendix \ref{AppAdS}), this algebra corresponds to $\mathfrak{sl}(2,\mathbb{R})\times \mathbb{R}$, with its relations given in \eqref{sl2R}. The non-degenerate invariant bilinear form is given \eqref{invtenZ} which in the conformal basis can be written as
\be
 \langle \tilde{\mathcal{D}}, \tilde{\mathcal{D}}\rangle= \gamma_0\,, \qquad  \langle \tilde{\mathcal{H}}, \tilde{\mathcal{K}}\rangle= -2\gamma_0, \qquad \langle \tilde{\mathcal{ Y}},\tilde{\mathcal{ Y}}\rangle =\gamma_1\,.
\ee
where $\tilde{\mathcal Y} \equiv \tilde Y$. Now, let us consider the local parametrisation of a group element $\tilde U$ associated to \eqref{sl2R} as follows
 \be\label{groupelementqua}
  \tilde{U} \, = \, e^{\tilde s \tilde{\mathcal{ Y}}} \,  e^{\tilde \rho \tilde{\mathcal{ H}}} \,  e^{\tilde y \tilde{\mathcal{ K}}} \, e^{\tilde u \tilde{\mathcal{D}}}  \,,
 \ee
where $\tilde s  , \tilde \rho, \tilde y ,$  and $\tilde u$ are the group manifold coordinates. Using the matrix representation \eqref{sl2Rirrep}, and considering $\tilde{\mathcal {Y}} = \mathbb{1}_{2\times 2}$, the group element in \eqref{groupelementqua} can be written as
\be
\tilde{U} \, =\,  \left(
\begin{array}{cc}
 \displaystyle e^{\tilde{s}+\frac{\tilde{u}}{2}} &\quad  -\tilde{y} \, e^{\tilde{s}-\frac{\tilde{u}}{2}} \\[1em]
 \tilde{\rho }\, e^{\tilde{s}+\frac{\tilde{u}}{2}} &\quad  e^{\tilde{s}-\frac{\tilde{u}}{2}} \left(1-\tilde{\rho } \,\tilde{y}\right) \\
\end{array}
\right)\,.
\ee
The MC one-form $\tilde \Omega= \tilde{U}^{-1} d \tilde{U}$ is given by
\be\label{MCformsl2u1}
\tilde \Omega \, = \,  \tilde{ \Omega}_{\tilde{\mathcal {H}}  }\, \tilde{\mathcal {H}} + \tilde \Omega_{\tilde{\mathcal {D}}  } \,\tilde{\mathcal {D}} + \tilde \Omega_{\tilde{\mathcal {K}}  }\, \tilde{\mathcal {K}} + \tilde \Omega_{\tilde{\mathcal {Y}}  }\, \tilde{\mathcal {Y}}\,, 
\ee
where the one-forms have the form 
\begin{multicols}{2}
\begin{subequations}\label{MCformsGalajinskylikequadratic}
\setlength{\abovedisplayskip}{-13pt}
\allowdisplaybreaks
\begin{align}
\tilde \Omega_{\tilde{\mathcal{H}}} & = e^{\tilde u} \, d \tilde \rho \,,\\[.2truecm]
 \tilde \Omega_{\tilde{\mathcal{K}}} & = e^{-\tilde{u}} \left(\tilde{y}^2\, d\tilde{\rho} +d\tilde{y}\right)   \,,\\[.2truecm]
\tilde \Omega_{\tilde{\mathcal{D}}} &= d \tilde u +2\tilde y \, d \tilde \rho  \,,\\[.4truecm]
 \tilde \Omega_{\tilde{\mathcal{Y}}} &= d\tilde s\,.
\end{align}
\end{subequations}
\end{multicols}\noindent
The MC one-forms can be used to construct the boundary action \eqref{boundaryaction3}
\be\label{conformalrelativisticaction}
S[\tilde \rho,\tilde y, \tilde u,\tilde s] \,= -\frac{k}{2}\int  d t \, \left(\langle\tilde \Omega,\tilde \Omega\rangle\right)^{\star}  \, =-\frac{k}{2} \int d t \,\left( \gamma_0 \,\left(4 \tilde{\rho }' \left(\tilde{y} \,\tilde{u}'-\tilde{y}'\right)+\tilde{u}'^2\right) +\gamma_1\, \tilde{s}'^2 \right)  \,,
\ee
which is invariant under the following global symmetries
\bseq\label{relativisticglobalsymmetries}
\ba 
\delta \tilde \rho & = & \tilde \beta -\tilde \alpha \tilde \rho +\tilde \gamma^2 \tilde \rho^2 \,,\\
\delta \tilde u & = & \tilde \alpha -2 \tilde \gamma\tilde \rho \,, \\
\delta \tilde y &=&\tilde \gamma +\tilde y (\tilde \alpha -2\tilde \gamma \tilde \rho) \,,\\
\delta \tilde s & = &\tilde \theta\,,
\ea 
\eseq
associated to the infinitesimal symmetry parameter 
\be
 \tilde \epsilon \, =\, \tilde \alpha \, \tilde{\mathcal{H}}+ \tilde \beta \, \tilde{\mathcal{D}}+ \tilde \gamma \, \tilde{\mathcal{K}} + \tilde{\theta}\, \tilde{\mathcal{Y}} \, = \, \begin{pmatrix}
\frac{1}{2}   \tilde{\beta} +\tilde \theta &-   \tilde{\gamma} \\[1em]
  \tilde{\alpha}& -\frac{1}{2}   \tilde{\beta}+\tilde \theta
 \end{pmatrix}\,.
\ee

In the following, we consider the following inverse Higgs constraints \cite{Galajinsky:2019lak}
\be\label{relativisticconstraints}
 \tilde{ \Omega}_{\tilde{\mathcal {H}}  }=\mu \, \tilde{ \Omega}_{\tilde{\mathcal {Y}}  } \,, \qquad \tilde{ \Omega}_{\tilde{\mathcal {D}}  }=-2\nu \, \tilde{ \Omega}_{\tilde{\mathcal {Y}}  }\,,
\ee
where $\mu$ and $\nu$ are arbitrary constants.  We can express $\tilde y$ and $\tilde u$ in terms of the independent Goldstone fields $\tilde \rho$ and $\tilde s$ as
\be\label{relativisticconstraints2}
\tilde y \, =\, - \frac{2\,\nu \,\tilde{s}^\prime+\tilde{u}^\prime}{2\tilde{\rho}^\prime}\,,
\qquad
\tilde u= {\rm Log}\left(\frac{\mu \, \tilde{s}^\prime}{\tilde{\rho}^\prime}\right)\,.
\ee
In this case, the MC one-form \eqref{MCformsl2u1} reduces to
\be\label{bcondu1}
\tilde{a}\, \equiv \, \tilde \Omega^{\star}\bigg\rvert_{\scriptscriptstyle IHM}  \, = \, \mu\, \tilde{s}' \, \tilde{\mathcal{H}}+ \tilde{s}' \, \tilde{\mathcal{Y}}-2 \nu \, \tilde{s}' \tilde{\mathcal{D}}+\frac{1}{2\mu \tilde{s}'} \left( \text{Sch}(\tilde \rho,t) - \text{Sch}(\tilde s,t) + \nu^2 \tilde{s}'^2\right) \tilde{\mathcal{K}} \,.
\ee
and therefore the action \eqref{conformalrelativisticaction} is reduced to
\be\label{Two-Schwarzians}
S[\tilde \rho, \tilde s]  \, = \,-\frac{k}{2} \int d t \left( -2\gamma_0 \left( \text{Sch}(\tilde \rho,t) - \text{Sch}(\tilde s,t)\right)+\gamma_1\,\tilde{s}'^{2}\right)  \,.
\ee
Note that, if we impose the condition $\tilde{s} -t=0$ in \eqref{Two-Schwarzians}, we recover the Schwarzian action \eqref{OneSchw}.

To construct the gauge field in the bulk, we can introduce a radial dependence by performing a gauge transformation \eqref{Aanda} with the group element
\be
b=\begin{pmatrix}
e^{r/2} &0 \\[1em]
0 &e^{-r/2}  
\end{pmatrix}\,.
\ee
Using the matrix representation \eqref{sl2Rirrep}, this leads to the following form for the connection \eqref{relnonrelconnection0}

\be
A\,=\,
\begin{pmatrix}
\frac{dr}{2}& 0 \\[1em]
0&-\frac{dr}{2}
\end{pmatrix}
+dt
\begin{pmatrix}
(1-\nu) \, \tilde{s}^\prime &\qquad -e^{-r} \, \left[\frac{1}{2\mu \tilde{s}^\prime}({\rm Sch}(\tilde\rho, t)-{\rm Sch}(\tilde s,t))+ \frac{\nu^2}{\mu}  \tilde{s}^\prime  \right]\\[1em]
\mu \,e^{r} \,\tilde{s}^\prime  &\qquad (1+\nu)\, \tilde{s}^\prime
\end{pmatrix}\,,
\ee
from which we can read off the zweibein and spin connection. In the AdS$_2$ case they read
\be
\bal
E^0&=\left(\mu s^\prime e^r-  \left[\frac{1}{2\mu \tilde{s}^\prime}({\rm Sch}(\tilde\rho, t)-{\rm Sch}(\tilde s,t))+ \frac{\nu^2}{\mu}  \tilde{s}^\prime  \right]e^{-r}\right)dt\,,\\
E^1&=dr\,,\\
\Omega& =\left(\mu s^\prime e^r+  \left[\frac{1}{2\mu \tilde{s}^\prime}({\rm Sch}(\tilde\rho, t)-{\rm Sch}(\tilde s,t))+ \frac{\nu^2}{\mu}  \tilde{s}^\prime  \right]e^{-r}\right)dt \,,\\
\eal
\ee
whereas, for the dS$_2$ case we find
\be\label{esomegasadsR}
\bal
E^0 &=dt\,,\\
E^1&=\left(\mu s^\prime e^t-  \left[\frac{1}{2\mu \tilde{s}^\prime}({\rm Sch}(\tilde\rho, r)-{\rm Sch}(\tilde s,r))+ \frac{\nu^2}{\mu}  \tilde{s}^\prime  \right]e^{-t}\right)dr\,,\\
\Omega& =\left(\mu s^\prime e^t+  \left[\frac{1}{2\mu \tilde{s}^\prime}({\rm Sch}(\tilde\rho, r)-{\rm Sch}(\tilde s,r))+ \frac{\nu^2}{\mu}  \tilde{s}^\prime  \right]e^{-t}\right)dr \,,\\
\eal
\ee
As done in \eqref{esomegads}, we have interchanged the coordinates $t$ and $r$ in the dS$_2$ case \eqref{esomegasadsR}, since the boundary of the de-Sitter space is at time-like infinity (in this case the prime denotes derivative with respect to the radial coordinate). From these expressions we can construct the space-time metrics
\ba
ds^2_{\scriptstyle AdS} &= & 
- \Bigg(\mu^2 s^{\prime 2} e^{2r} -
 \left[({\rm Sch}(\tilde\rho, t)-{\rm Sch}(\tilde s,t))+ 2\nu^2 \tilde{s}^{\prime 2}  \right]\\
 &&
\qquad+e^{-2r} \left[\frac{1}{2\mu \tilde{s}^\prime}({\rm Sch}(\tilde\rho, t)-{\rm Sch}(\tilde s,t))+ \frac{\nu^2}{\mu}  \tilde{s}^\prime  \right]^2\Bigg)dt^2+dr^2\,, \\
ds^2_{\scriptstyle dS} &= &  \Bigg(\mu^2 s^{\prime 2} e^{2r} -
 \left[({\rm Sch}(\tilde\rho, t)-{\rm Sch}(\tilde s,t))+ 2\nu^2 \tilde{s}^{\prime 2}  \right]\\
 &&
\qquad+e^{-2r} \left[\frac{1}{2\mu \tilde{s}^\prime}({\rm Sch}(\tilde\rho, t)-{\rm Sch}(\tilde s,t))+ \frac{\nu^2}{\mu}  \tilde{s}^\prime  \right]^2\Bigg)dt^2-dr^2\,.
\ea

Analogously to the previous section, one can ask what are the asymptotic symmetries associated with the theory defined by \eqref{Two-Schwarzians}. For simplicity let us consider the case $\nu =0$, and rewrite the boundary field \eqref{bcondu1} as follows
\be\label{asl2rxr}
\tilde a \, = \,\left(\mu \, \mathcal{T}  \, \mathcal{\tilde{H}} + \frac{1}{\mu} \,\frac{ \mathcal{L}}{\mathcal{T} } \, \mathcal{\tilde{K}} + \mathcal{T} \, \mathcal{\tilde{Y}} \right) \, dt\,, 
\ee
where we have defined the functions
\be
\mathcal{T}(t) \, \equiv \, \tilde{s}' \,, \qquad \mathcal{L}(t)\, \equiv \,  \frac{1}{2} \left( \text{Sch}(\tilde \rho,t) - \text{Sch}(\tilde s,t) \right) \,.
\ee
 The asymptotic symmetries correspond to the set of gauge transformations \eqref{tgpres} preserving 
 the functional form of \eqref{asl2rxr}. Considering a gauge parameter of the form
\be\label{gaugeparameterSL2Ru1}
\lambda  \, =\,\lambda_0  \, \tilde{\mathcal{H}} + \lambda_1  \, \tilde{\mathcal{K}} +  \lambda_2  \, \tilde{\mathcal{D}}+\lambda_3  \tilde{\mathcal{Y}}   \,.
\ee
one can solve \eqref{tgpres} to find
\bseq \label{gparametersl2Ru1}
\ba
 \lambda_0 & = &\mu \mathcal{T}\epsilon\,,\\
 \lambda_1 & = &  \frac{1}{\mu\mathcal{T} }\bigg[\frac{1}{2}\left(\epsilon^\prime + \frac{\mathcal{T}^\prime}{\mathcal{T}}\epsilon - \chi^\prime- \frac{\mathcal{T}^\prime}{\mathcal{T}} \chi\right)^\prime+ \mathcal{L}\epsilon\bigg]\,, \\
  \lambda_2 & = & -\epsilon^\prime - \frac{\mathcal{T}^\prime}{\mathcal{T}} \epsilon + \chi^\prime +  \frac{\mathcal{T}^\prime}{\mathcal{T}} \chi\,,\\
 \lambda_3 & = &  \mathcal{T}\chi \,.
\ea
\eseq
The functions $\mathcal{T}$ and $\mathcal L$ transform under the asymptotic symmetries as
\ba\label{variationsl2xR}
\delta \mathcal L &=  & \epsilon \, \mathcal L^\prime +2 \mathcal L\, \epsilon^\prime -  \frac{\mathcal{T}^\prime}{2\mathcal{T}}\,\left(\epsilon^\prime + \frac{\mathcal{T}^\prime}{\mathcal{T}}\epsilon - \chi^\prime- \frac{\mathcal{T}^\prime}{\mathcal{T}} \chi\right)^\prime+  \frac{1}{2}\left(\epsilon^\prime + \frac{\mathcal{T}^\prime}{\mathcal{T}}\epsilon - \chi^\prime- \frac{\mathcal{T}^\prime}{\mathcal{T}} \chi\right)^{\prime\prime}\,, \\
\delta \mathcal T & =& \chi^\prime \, \mathcal T +\chi\, \mathcal{T}^\prime\,. 
\ea
Using  the explicit form of the gauge parameters, the charge can be integrated to give
\be\label{chargesl2Ru1}
  Q [\epsilon, \chi] \, = L[\epsilon] + T[\chi]  \,,
\ee
where we have defined
\ba
L[\epsilon] & =& k \gamma_0  \, \int \epsilon  \left(  2\mathcal L + \left(\frac{\mathcal T^\prime}{\mathcal T}\right)^\prime
 -\frac{1}{2}\left(\frac{\mathcal T^\prime}{\mathcal T}\right)^2  \right) \, d\tau_{\scriptscriptstyle E} \,, \\
T[\chi]  & =& - \frac{ k }{2} \, \int \chi  \left(\gamma_1  \mathcal T^2+ \gamma_0\left(\frac{\mathcal T^\prime}{\mathcal T}\right)^\prime
 -\gamma_0\left(\frac{\mathcal T^\prime}{\mathcal T}\right)^2  \right) \, d\tau_{\scriptscriptstyle E} \,.
\ea
The Poisson algebra of these charges can be computed using \eqref{poissonbracket}, which yields two Virasoro algebras
\be\label{Poissonsl2xR}
\bal
\{  L[\epsilon_1], L[\epsilon_2] \} \,  &=-\, L[[\epsilon_1,\epsilon_2] ]-k\gamma_0 \int \, \epsilon_1^{\prime}  \,\epsilon_2^{\prime\prime} d \tau_{\scriptscriptstyle E}\,,\\
\{  T[\chi_1], T[\chi_2] \} \,  &=\, -T[[\chi_1,\chi_2] ]+k\gamma_0 \int  \,\chi_1^{\prime}\, \chi_2^{\prime\prime}d \tau_{\scriptscriptstyle E}\,.
\eal
\ee
Introducing the Fourier modes 
\be
\mathcal{L}_m = L[e^{im \tau_{\scriptscriptstyle E}}]  \,, \qquad \mathcal{T}_m =   T[e^{im \tau_{\scriptscriptstyle E}}] \,,
\ee
and changing to Euclidean time in the variations \eqref{variationsl2xR}, the corresponding Poisson algebra leads to two Virasoro algebras
\be
\bal
 \{ \mathcal L_m, \mathcal L_n \}  = & (m-n) \, \mathcal{L}_{n+m} -2\pi k \gamma_0\,m^3 \, \delta_{m+n,0}\,, \\
 \{ \mathcal T_m, \mathcal T_n \}  = & (m-n) \, \mathcal{T}_{n+m} +2\pi k \gamma_0\,m^3 \, \delta_{m+n,0}\,.
\eal
\ee
The central terms have the same form but with opposite sign, which is compatible with the form of the double Schwarzian boundary action \eqref{Two-Schwarzians}.

Finally, it is worth to mention that from the constraint $\tilde{ \Omega}_{\tilde{\mathcal {H}}  }\,= \,\mu \, \tilde{ \Omega}_{\tilde{\mathcal {Y}}  }$ we can express $\tilde{\rho}^\prime$ in terms of $\tilde{s}^\prime$ and $\tilde u$,
\be
\rho^\prime=e^{-u} s^\prime\,,
\ee
resulting in the action
\ba \label{relativisticmodelboundary}
  S[\tilde u, \tilde s]  & =  & \int d t \, \left(\gamma_0\left(\tilde{u}'^{2}+2 \tilde{u}^{\prime \prime }-\frac{2 \tilde{s}^{\prime \prime} \tilde{u}^{\prime}}{\tilde{s}^{\prime}} \right)+\gamma_1\, \tilde{s}'^{2}\right) \,.
\ea 
In the next section, we will show that this last model admits a finite NR limit on its fields, which leads to a NR conformal model invariant under a NR conformal symmetry isomorphic to the Extended Newton-Hooke algebra.

\section{Non-relativistic boundary theory}\label{NRschsection}
In this section, we construct the boundary theory associated to the NR JT gravity and compute the the asymptotic symmetries and its associated conserved charges. We show how these results can be extended to the Carrollian case.

\subsection{Non-relativistic limit of the $SL(2,\mathbb R)\times \mathbb R$ boundary action}\label{NRLimit-Boundarytheory}

Here, we derive a NR limit of the relativistic model in \eqref{relativisticmodelboundary}. In order to do that, we consider the following redefinition of the $SL(2,\mathbb R) \times \mathbb R$ generators
\begin{multicols}{2}
\begin{subequations}\label{transfConfBasis} 
\setlength{\abovedisplayskip}{-13pt}
\allowdisplaybreaks
\begin{align}
 \tilde{\mathcal{ D}} & = \frac{1}{2}\mathcal  D +\frac{1}{\varepsilon^2} \mathcal  Z \,,   \,\\[.1truecm]
 \tilde{\mathcal{ Y}}  & =  \frac{1}{2} \mathcal  D - \frac{1}{\varepsilon^2}  \mathcal  Z\,,  \,\\[.1truecm]
\tilde{\mathcal{H}} & = \frac{1}{\varepsilon} \, \mathcal  H \,,\\[.1truecm] 
\tilde{\mathcal{ K}} & = \frac{1}{\varepsilon} \, \mathcal  K \,,
\end{align}
\end{subequations}
\end{multicols}\noindent
whose inverse reads
\begin{multicols}{2}
\begin{subequations} \label{transfConfBasis2} 
\setlength{\abovedisplayskip}{-13pt}
\allowdisplaybreaks
\begin{align}
 \mathcal  D & = \tilde{\mathcal{D}} + \tilde{\mathcal{ Y}}  \,,\\[.1truecm]
\mathcal  Z  & =   \frac{\varepsilon^2}{2} (\tilde{\mathcal{ D}} - \tilde{\mathcal{ Y}})    \,,\\[.1truecm]
 \mathcal H & = \varepsilon\, \tilde{\mathcal{ H}} \,,\\[.4truecm] 
 \mathcal K & = \varepsilon \,\tilde{\mathcal{ K}} \,.
\end{align}
\end{subequations}
\end{multicols}\noindent
Taking the limit $\varepsilon \rightarrow 0$, we get
\be\label{NRcb}
[\mathcal H,\mathcal  D] = \mathcal H \,, \qquad [\mathcal K,\mathcal D] = -\mathcal K \,, \qquad [\mathcal H,\mathcal K] = 2\mathcal Z \,,
\ee
which we name {\it Extended Galilean conformal  (EGC) algebra} in 1D. This algebra is isomorphic to the NH$^{+}$ algebra. In fact, using the change of basis
\be\label{changenh+}
\mathcal D= \ell\, H \,,\hskip.8truecm
\mathcal H= \ell \,P+G \,,\hskip.8truecm
\mathcal K= \ell\, P -G \,,\hskip.8truecm
\mathcal Z=\ell \,M\,,
\ee
we find the commutation relations \eqref{newton-hookealgebra} with cosmological constant $\Lambda=1/\ell^2$
\ba \label{newton-hookealgebra+}
\left[ G , H\right]  = P, \quad \quad  \left[ G,  P\right]  = M , \quad \quad  \left[ H, P\right]  =  -\frac{1}{\ell^2} G\,.
\ea
Using the transformation \eqref{transfConfBasis} and its inverse \eqref{transfConfBasis2}, one can relate relativistic Goldstone fields in terms of NR ones as
\begin{multicols}{2}
\begin{subequations} 
\setlength{\abovedisplayskip}{-13pt}
\allowdisplaybreaks
\begin{align}\label{NRGoldstone1}
 \tilde u & = u + \frac{\varepsilon^2}{2 }  s \,,\\[.1truecm]
 \tilde \rho  & =  \varepsilon \, \rho    \,,\\[.1truecm]
 \tilde y & = \varepsilon \, y \,,\\[.1truecm] 
 \tilde s & = u - \frac{\varepsilon^2}{2 }  s \,.\label{NRGoldstone2}
\end{align}
\end{subequations}
\end{multicols}\noindent
Therefore, applying the field transformations \eqref{NRGoldstone1} and \eqref{NRGoldstone2} on the action \eqref{relativisticmodelboundary}, we find
\be
S[u,s] = 2 \, \varepsilon^2 \,\gamma_0 \, \int dt\,\left( s''+s' \left(u'-\frac{u''}{u'}\right)\right) + \mathcal{O} \left( \varepsilon^4\right)\,,
\ee
where we have set $\gamma_1=-\gamma_0$ in order to cancel a divergent term. Now, defining $\gamma_0 :=  \alpha_0/\varepsilon^2 $, in the limit $\varepsilon \rightarrow 0$, the NR model becomes
\be\label{NRSchwarzian}
S[u,s] = 2\, \alpha_0\int dt \, \left( s''+s' \left(u'-\frac{u''}{u'}\right)\right) \,.
\ee
We shall refer to this action as the {\it Non-Relativistic Schwarzian}. Note that this boundary action appears at order in $\varepsilon^2$ in the same way that the NR gravity theory on the bulk does in \eqref{NRaction0}. It is important to remark that if we perform the NR limit of the boundary action when expressed in terms of Schwarzians \eqref{Two-Schwarzians}, the divergent term is not cancelled.

In order to obtain the NH$^-$ algebra, we consider the following redefinition of the \aslu generators
\begin{multicols}{2}
\begin{subequations}\label{contractionsl22}
\setlength{\abovedisplayskip}{-13pt}
\allowdisplaybreaks
\begin{align}
\tilde{\mathcal H}  & = \frac{1}{2}\mathcal D 
+\frac{1}{2\varepsilon}(\mathcal H -\mathcal K)
+\frac{1}{\varepsilon^2} \mathcal Z \,,   \,\\[.1truecm]
\tilde{\mathcal K}  & = \frac{1}{2}\mathcal D 
-\frac{1}{2\varepsilon}(\mathcal H -\mathcal K)
+\frac{1}{\varepsilon^2} \mathcal Z \,,   \,\\[.1truecm]
\tilde{\mathcal D} & = \frac{1}{2\varepsilon}(\mathcal H+ \mathcal K )\,,  \,\\[.1truecm]
\tilde{\mathcal Y}  & =  \frac{1}{2}\mathcal D - \frac{1}{\varepsilon^2} \mathcal Z\,.
\end{align}
\end{subequations}
\end{multicols}\noindent
Inverting these relations leads to
\begin{multicols}{2}
\begin{subequations} 
\setlength{\abovedisplayskip}{-13pt}
\allowdisplaybreaks
\begin{align}
 \mathcal H & =  \varepsilon  \left[\tilde{\mathcal D} +\frac{1}{2}(\tilde{\mathcal H} -\tilde{\mathcal K} ) \right]\,,\\[.1truecm] 
 \mathcal K & =   \varepsilon  \left[\tilde{\mathcal D} -\frac{1}{2}(\tilde{\mathcal H} -\tilde{\mathcal K} ) \right] \,,\\[.1truecm]
 \mathcal D & = \frac{1}{2}(\tilde{\mathcal H} +\tilde{\mathcal K} ) + \tilde{\mathcal Y}   \,,\\[.1truecm]
 \mathcal Z  & =   \frac{\varepsilon^2}{2 } \left[\frac{1}{2}(\tilde{\mathcal H} +\tilde{\mathcal K} ) - \tilde{\mathcal Y}    \right] \,.
\end{align}
\end{subequations}
\end{multicols}\noindent
In the limit $\varepsilon\rightarrow0$, we find the following NR contraction of \aslu
\be\label{alt}
[\hat{\mathcal H}, \hat{\mathcal D}]=\hat{\mathcal K}\,,\qquad
[\hat{\mathcal K}, \hat{\mathcal D}]=-\hat{\mathcal H}\,,\qquad
[\hat{\mathcal H}, \hat{\mathcal K}]=2\hat{\mathcal Z}\,,\qquad
\ee
which will be referred to as {\it Twisted Extended Galilean Conformal algebra} in 1D. By defining the following change of basis
\be\label{changenh-}
\hat{\mathcal D}= \ell H \,,\hskip.8truecm
\hat{\mathcal H}= \ell P+G \,,\hskip.8truecm
\hat{\mathcal K}= \ell P -G \,,\hskip.8truecm
\hat{\mathcal Z}=\ell M\,,
\ee
we find the NH$_2^-$ algebra \eqref{newton-hookealgebra} with $\Lambda=-1/\ell^2$
\ba \label{Enewton-hookealgebra-}
\left[ G , H\right] = P, \quad \quad  \left[ G,  P\right]  = M , \quad \quad  \left[ H, P\right]  = \frac{1}{\ell^2} G\,.
\ea
It is important to note that the NR symmetries \eqref{NRcb} and \eqref{alt} are related by the complex change of basis 
\begin{multicols}{2}\label{changeNH+toNH-}
\begin{subequations} 
\setlength{\abovedisplayskip}{-9pt}
\allowdisplaybreaks
\begin{align}
 \hat{\mathcal{D}} & =- i \, \mathcal{D}    \,,\\[.4truecm]
\hat{\mathcal{Z}}  & =2i\,  \mathcal{Z}  \,,\\[.4truecm]
   \hat{\mathcal{H}} & =\mathcal{H}+\mathcal{K}   \,,  \\[.4truecm]
\hat{\mathcal{K}}&=-i\left( \mathcal{H} -  \mathcal{K}\right)\,.
\end{align}
\end{subequations}
\end{multicols}\noindent
Therefore, one can consider the analogue change of variables between the relativistic Goldstone fields $\tilde s\,, \tilde u$ and the ones of the Extended NH$^{-}$ algebra $\hat s \,, \hat u$, i.e.
\be
\tilde u =-i\, \hat u + \frac{i \, \varepsilon^2}{2}\hat{s} \,, \qquad \qquad  \tilde s  =-i\, \hat u - \frac{i \, \varepsilon^2}{2}\hat{s} \,. 
\ee 
Applying these field transformations into \eqref{relativisticmodelboundary}, we get  
\be
S[\hat u , \hat s] \, =\,2 i \varepsilon^2 \, \gamma_{0} \int dt\, \left( \hat{s}'' - \hat{s}' \left(   i\, \hat{u}' +\frac{\hat{u}''}{\hat{u}'}  \right) \right) + \mathcal{O} \left( \varepsilon^4\right)  \,,
\ee
again setting $\gamma_1=-\gamma_0$, and taking the definition $\gamma_0 :=  \alpha_0/\varepsilon^2 $, in the limit $\varepsilon \rightarrow 0$, we get the complexified version of NR Schwarzian \eqref{NRSchwarzian}, namely 
\be\label{ComplexNRSchwarzian}
S[\hat u , \hat s] \, = \, 2 i  \, \alpha_{0} \int dt\, \left( \hat{s}'' - \hat{s}' \left(   i\, \hat{u}' +\frac{\hat{u}''}{\hat{u}'} \right) \right)   \,.
\ee
This action is closely related to the one obtained in \cite{Afshar:2019axx}. In the following, we will derive these same actions using an IHM.

\subsection{Extended Newton-Hooke$_2^+$ case} \label{secextnh+}
Following the steps outlined in Section \ref{schsection}, we construct the MC one-form associated to the Extended Galilean conformal algebra \eqref{NRcb} by considering a local parametrisation of a group element $U$ as
\be\label{coset+}
U  \,= \,  e^{s \mathcal{Z}} \,   e^{\rho \mathcal{H}} \,  e^{y \mathcal{K}} \,  e^{u \mathcal{D}} \,,
\ee 
where $ s,\rho,y$, and $u$ correspond to the set of NR coordinates of the group manifold. We consider a faithful $3\times3$ matrix representation of \eqref{NRcb} given by
\ba
&& \mathcal{Z} = \left(
\begin{array}{ccc}
 0 & 0 & \frac{1}{2} \\
 0 & 0 & 0 \\
 0 & 0 & 0 \\
\end{array}
\right) \,,
  \qquad  \mathcal{H} =    \left(
\begin{array}{ccc}
 0 & 1 & 0 \\
 0 & 0 & 0 \\
 0 & 0 & 0 \\
\end{array}
\right) \,, \qquad \mathcal{K}=\left(
\begin{array}{ccc}
 0 & 0 & 0 \\
 0 & 0 & 1 \\
 0 & 0 & 0 \\
\end{array}
\right) \,,  \qquad
 \mathcal{D} = \left(
\begin{array}{ccc}
 0 & 0 & 0 \\
 0 & 1 & 0 \\
 0 & 0 & 0 \\
\end{array}
\right) \,, 
\ea 
so that the group element $U$ reads
\ba
U & = & \left(
\begin{array}{ccc}
 1 &\quad  \rho\,  e^{u} &\quad  \displaystyle \frac{s}{2}+\rho \, y \\
 0 &\quad  e^{u} &\quad  y \\
 0 & \quad 0 &\quad  1 \\
\end{array}
\right)\,.
\ea
The left-invariant MC one-form is then given by
 \be 
 \Omega\, =\,U^{-1} d U \, = \, \Omega_{\mathcal{Z}} \, \mathcal{Z}+\Omega_{\mathcal{H}} \, \mathcal{H}+\Omega_{\mathcal{K}} \, \mathcal{K}+\Omega_{\mathcal{D}} \, \mathcal{D} \,,
 \ee
with components
\ba 
\Omega_{\mathcal Z}  =  ds +2\, y\, d \rho\,, \qquad \Omega_{\mathcal H}  =  e^{u} \, d \rho \,, \qquad 
\Omega_{\mathcal  K}  =  e^{-u} \, dy\,, \qquad 
\Omega_{\mathcal D}  =  d u \,. \qquad 
\ea
The invariant bilinear form in this case reads
\eqref{NRcb} 
\be\label{pairingNH+}
 \langle \mathcal{D}, \mathcal{D} \rangle=c_0 \,, \qquad \langle \mathcal{D}, \mathcal{Z} \rangle=c_1\,, \qquad  \langle \mathcal{H}, \mathcal{K} \rangle=-2\,c_1 \,,
\ee
where $c_0$ and $c_1$ are arbitrary constants. This bilinear form is non-degenerate for $c_1 \neq 0$. Using these expressions, the boundary action \eqref{boundaryaction3} takes the form
\be\label{NRboundarymoldel}
S[s,\rho,y,u]\,=\,  \int dt \, \left( c_0\, u'^{2} + 2c_1 \, \left(u' (s'+2y \rho') -2\rho' y'\right)   \right)  \,.
\ee
and is invariant under the following global symmetry transformations 
\ba\label{globalsymmetriesconventionalalgebra}
\delta s=  \theta-2  \gamma  \rho  \,, \qquad \delta \rho =  \beta - \alpha\,  \rho \,, \qquad \delta y=  \gamma + \alpha  y\,, \qquad \delta u= \alpha\,,
\ea 
where the infinitesimal parameters can be  grouped as
$\epsilon = \theta \, \mathcal{Z}+ \beta \, \mathcal{H}  +  \gamma \, \mathcal{K} +  \alpha \, \mathcal{D}$\,. 

In order to reduce the number of Goldstone fields, we will impose a particular IHM constraint. Let us consider the constraints 
\be
 \Omega_{\mathcal Z} = 0\,, \qquad  \Omega_{\mathcal H} =\alpha \,,
\ee 
with $\alpha$ a constant. The boundary theory \eqref{NRboundarymoldel} then reduces to 
\be \label{reducedaction+2}
 S[u,s]=  \int dt \,\left( c_0 \,u'^2+2 \,c_1 \left( s''+s' u' \right) \right)\,.
\ee
Note that, performing the change of variables
\be
u=v-{\rm Log} ( v^\prime)\,,
\ee
and setting $c_0=0$ in the bilinear form \eqref{pairingNH+}, we recover the action \eqref{NRSchwarzian}
\be \label{reducedaction+3}
 S[v,s]= 2 \,c_1 \int dt \,  \left( s''+s' \left( v^\prime-\frac{ v^{\prime\prime}}{v^\prime}\right)\right)  \,.
\ee
In this case the boundary connection takes the form
\be\label{anh+}
a=\Omega^*\bigg|_{\scriptscriptstyle IHM}= \bigg(\alpha {\mathcal {H}} +   \mathcal T(t) \, {\mathcal {D}} +\mathcal L (t) {\mathcal {K}}  \bigg)dt\,= \, \left(
\begin{array}{ccc}
 0 & \alpha  & 0 \\
 0 &\mathcal T (t)  &\mathcal L (t) \\
 0 & 0 & 0 \\
\end{array}
\right)
\ee
where we have defined the functions
\be
\mathcal T (t) \, \equiv \,  v^\prime-\frac{v^{\prime \prime }}{v^\prime} \,, \qquad  \mathcal L (t) \, \equiv \,  \frac{1}{2 \alpha v^\prime }\left(s^\prime \left(v^{\prime \prime }-v^{\prime 2}\right)-s^{\prime \prime } v^\prime\right)\,.
\ee
The gauge transformations preserving these boundary conditions are obtained by solving \eqref{tgpres}, which gives
\be
\bal
\lambda_{\mathcal Z}&= -2\alpha \chi \,,\\
\lambda_{\mathcal H}&= \alpha \sigma \,,\\
\lambda_{\mathcal K}&= \chi^\prime +\mathcal L \sigma \,,\\
\lambda_{\mathcal D}&=  \sigma^\prime +\mathcal T \sigma \,,
\eal
\ee
whereas the variations of $\mathcal L$ and $\mathcal T$
read
\ba\label{varLTnh+}
\delta \mathcal L(t) &=  & \sigma \, \mathcal L^\prime +2 \mathcal L\, \sigma^\prime + \mathcal T\, \chi^\prime- \chi^{\prime \prime }\,, \\
\delta \mathcal T(t) & =& \sigma^\prime \, \mathcal T +\sigma\, \mathcal T^\prime -\sigma^{\prime \prime} \,. 
\ea
The charge \eqref{deltaQgen} can be integrated, leading to
\be
Q[\sigma, \chi]= L[\sigma] + T[\chi]\,,
\ee 
where
\be
\bal
L[\sigma]&=2\alpha c_1 k\int  \sigma \,\mathcal L d  \tau_{\scriptscriptstyle E}\,,\\
T[\chi]&=2\alpha c_1  k\int \, \chi \,\mathcal Td  \tau_{\scriptscriptstyle E}\,.
\eal
\ee
The Poisson algebra \eqref{poissonbracket} in this case takes the form of the Twisted Warped Virasoro algebra \cite{Afshar:2019axx,Hofman:2014loa,Godet:2020xpk}, which after expanding in modes
\be
\mathcal L_m \equiv L[e^{im \tau_{\scriptscriptstyle E}}] \,, \qquad \mathcal T_m \equiv T[e^{im \tau_{\scriptscriptstyle E}}] \,,
\ee
has the form
\be
\bal\label{warpedV}
 \{ \mathcal L_m, \mathcal L_n \} = & (m-n) \, \mathcal{L}_{n+m} \,, \\
  \{ \mathcal L_m, \mathcal T_n \}  = & 4 \pi \alpha c_1 m \, \delta_{n+m,0}\,, \\
 \{ \mathcal T_m, \mathcal T_n \}  = &0\,.
\eal
\ee

We can introduce the dependence on a radial coordinate $r$ by applying a gauge transformation with an $r$-dependent group element of the form \cite{Afshar:2019axx}
\be
b=e^{-r \mathcal H}= \left(
\begin{array}{ccc}
 1 &-r & 0 \\
 0 &1 &0 \\
 0 & 0 &  1 \\
\end{array}
\right)\,.
\ee
This leads to 
\be
A=\left(-dr+(r \mathcal T(t) +\alpha) dt \right) \mathcal H + \left(
\mathcal L(t)  \,\mathcal K
+\mathcal T(t)\,  \mathcal D
+2r \mathcal L(t)\, \mathcal Z\right)dt\,.
\ee
Using the change of basis \eqref{changenh+} we can find the asymptotic form of the NR gravitational gauge fields
  \be
 \bal
 \tau \bigg|_{\partial \mathcal{M}} \, =\, &\ell\, \mathcal T dt\,,\\
 e\bigg|_{\partial \mathcal{M} } \, = \, & \ell\left(-dr+(r\mathcal T  +\alpha +\mathcal L )dt\right)\,,\\
  \omega \bigg|_{\partial \mathcal{M}} \,= \, & -dr+(r\mathcal T +\alpha  -\mathcal L) dt\,,\\
 m \bigg|_{\partial \mathcal{M}} \, = \,& 2\ell \,r\, \mathcal L dt \,.
\eal 
 \ee
which satisfy the field equations \eqref{fieldeqS1R}. In this case, the Newton-Cartan geometric structure is described by the asymptotic spatial metric
\be
h_{\mu\nu}= \ell^2\bigg[ \left( r \mathcal T+\alpha  +\mathcal L\right)^2 dt^2 -2\left( r\mathcal T + \alpha +\mathcal L\right) dr dt+  dr^2\bigg]\,,
\ee
which is degenerate with null vector 
\be
\tau^\mu \, = \,\frac{1}{\ell \mathcal T}\left(1\,, \,\mathcal L +\alpha + r \mathcal T\right) \,.
\ee

\subsection{Extended Newton-Hooke$_2^{-}$ case}
Now we consider the boundary action associated to the Twisted Extended Conformal algebra \eqref{alt}, isomorphic to the NH$_2^-$ symmetry. The NR analysis in the case of negative cosmological constant can be done in the same way as the previous case by using the the Extended Galilean conformal algebra \eqref{NRcb} and the complex change of basis \eqref{changeNH+toNH-}. This relates the coordinates of the coset \eqref{coset+} to new coordinates $\{ \hat{s}, \hat{\rho},\hat{y},\hat{u}\}$ associated to the algebra \eqref{alt},
\begin{multicols}{2}
\begin{subequations} 
\setlength{\abovedisplayskip}{-13pt}
\allowdisplaybreaks
\begin{align}
\rho &  =\frac{1}{\sqrt{2}}  \left( \hat{ \rho} -i\, \hat{y} \right)\,,\\[.2truecm]
y & =\frac{1}{\sqrt{2}}  \left( \hat{ \rho} +i\, \hat{y} \right)  \,,\\[.1truecm]
u &=-i \, \hat{u}\,,  \\[.5truecm]
s&=i \, \hat{s}\,.
\end{align}
\end{subequations}
\end{multicols}\noindent
Applying these transformations directly to the reduced action \eqref{NRboundarymoldel}, we find
\be\label{complexifiedNRaction}
S[\hat u , \hat s] = i \int d\tau\, \left( i \,c_0 \, \hat{u}'^2 +2 \,c_1 \left( \hat{s}'' - \hat{s}' \left(   i\, \hat{u}' +\frac{\hat{u}''}{\hat{u}'} \right) \right) \right)   \,.
\ee
This action \eqref{complexifiedNRaction} is invariant under the global symmetries 
\ba 
 \delta \hat{s}  = \tilde{\theta} + i (\hat{\beta}  +i \hat{\gamma}  ) (\hat{\rho} -i\, \hat{y})\,, \qquad
 \delta \hat{\rho}  = \hat \beta   + \hat{\alpha} \hat{y}\,,  \qquad 
 \delta \hat{y} =  \hat{\gamma} - \hat{\alpha} \hat{\rho} \,, \qquad \delta \hat  u = \hat{\alpha}\,,
\ea
with symmetry parameter $\hat{ \epsilon} \, =\, \hat{ \theta} \, \hat{\mathcal{Z}}+ \hat{\beta} \, \hat{\mathcal{H}} + \hat{\gamma} \, \hat{\mathcal{K}} + \hat{\alpha} \, \hat{\mathcal{D}}\,.$
The action \eqref{complexifiedNRaction} matches the model found in \cite{Afshar:2019axx} as the flat limit of the Schwarzian action. 

To perform the asymptotic analysis, we can use \eqref{anh+} and the change of basis \eqref{changeNH+toNH-} to write the boundary connection as
\be
a\equiv \Omega^*\bigg|_{\scriptscriptstyle IHM} \, = \, \bigg(
\frac{1}{2}\left(\alpha + \mathcal{L}(t)\right) \hat{\mathcal {H}} + i  \mathcal{T}(t) \, \hat{\mathcal{D}} +\frac{i}{2}\left(\alpha - \mathcal{L}(t)\right) \hat{\mathcal{K}}  \bigg)\,dt\,,
\ee
Then, the same analysis as the previous section leads to the variations \eqref{varLTnh+} for the functions $\mathcal{L}(t)$ and $\mathcal{T}(t)$ and to the asymptotic symmetry \eqref{warpedV}.


\subsection{Extended Carroll (A)dS$_2$ case}

Due to the duality \eqref{dualities}, the Extended Conformal Galilean algebra \eqref{NRcb} and its twisted version \eqref{alt} are isomorphic to Carrollian symmetries. Indeed starting from \eqref{NRcb} and using the change of basis 
\be\label{basisCarrollAdS}
\mathcal D= \ell P \,,\hskip.8truecm
\mathcal H= \ell H+G \,,\hskip.8truecm
\mathcal K= \ell H -G \,,\hskip.8truecm
\mathcal Z=\ell M\,,
\ee
leads to the Carroll AdS$_2$ algebra
\ba \label{carrolladsalg}
\left[ G , P\right] = H, \quad \quad  \left[ G,  H\right]  = M , \quad \quad  \left[ H, P\right]  = \frac{1}{\ell^2} G\,.
\ea
Similarly, applying the change of basis
\be\label{changetoeucnh}
\mathcal D= \ell P \,,\hskip.8truecm
\mathcal H= \ell H+G \,,\hskip.8truecm
\mathcal K= \ell H -G \,,\hskip.8truecm
\mathcal Z=\ell M\,,
\ee
to the algebra \eqref{alt} yields the Carroll dS$_2$ algebra
\ba \label{carrolddsalgebra}
\left[ G , P\right] = H, \quad \quad  \left[ G,  H\right]  = M , \quad \quad  \left[ H, P\right]  = -\frac{1}{\ell^2} G\,.
\ea
This means that the boundary actions \eqref{reducedaction+2} and \eqref{complexifiedNRaction} can also be interpreted as a boundary action of Carrollian JT gravity. Thus, the asymptotic symmetry analysis done in Section \ref{secextnh+} also holds here and leads to the Warped Virasoro symmetry \eqref{warpedV}.

Similarly to the NR case, the change of variables \eqref{basisCarrollAdS} for the Carroll AdS$_2$ case can be used to define a Carrollian geometry. Indeed, the asymptotic form of the Carrollian gravitational gauge fields read
  \be\label{Carrollstructure}
 \bal
 e \bigg|_{\partial \mathcal{M}} \, =\, &\ell \, \mathcal T dt\,,\\
 \tau \bigg|_{\partial \mathcal{M} } \, = \, & \ell\left(-dr+(r\mathcal T  +\alpha  +\mathcal L) dt\right)\,,\\
  \omega \bigg|_{\partial \mathcal{M}} \,= \, & -dr+(r\mathcal T  +\alpha  -\mathcal L )dt\,,\\
 m \bigg|_{\partial \mathcal{M}} \, = \,& 2\,\ell\, r \,\mathcal L dt \,.
\eal 
 \ee
In this case, the Carrollian geometric structure is described by the temporal metric
\be
\tau_{\mu\nu} \, \equiv \,   -\tau_\mu \tau_\nu \, = \, -\ell^2\bigg[ \left( r \mathcal T+\alpha  +\mathcal L\right)^2 dt^2 -2\left( r\mathcal T + \alpha +\mathcal L\right) dr dt+  dr^2\bigg]\,,
\ee
which is degenerate and has a null vector given by
\be
e^\mu \, = \,\frac{1}{\ell \mathcal T}\left(1\,, \,\mathcal L +\alpha + r \mathcal T\right) \,.
\ee
A similar construction for the Carroll dS$_2$ can be done by means of the change of variables \eqref{changetoeucnh}.

\section{Conclusions and Outlook}\label{outlook}
We have constructed the NR and 
Carrollian limits of JT gravity theory and the corresponding boundary actions.
The analysis in the bulk is done first by considering the NR 
and Carrollian limits of a BF action
with gauge algebra (A)dS$_2\times \mathbb R$. Next, we have
constructed BF theories with gauge algebras given by NH$^\pm$ and Carroll (A)dS$_2$. The Carrollian analysis is easily done using the duality \eqref{dualities} between NR and Carrollian symmetries. The he NR and Carrollian second order theories in the bulk are constructed by considering a suitable expansion of the metric tensor up to order $1/c^2$ and $c^2$, respectively. The actions are such that no divergent terms appear in the expansion if the Newton's constant is properly rescaled. This property is due to the fact that in two dimensions there are no divergent terms in the expansion of the Ricci scalar. This is a remarkably property that is in high contrast with its analogue in three and four space-time dimensions  \cite{DePietri:1994je,VandenBleeken:2017rij,Hansen:2020pqs,Bergshoeff:2019ctr}.

The boundary actions associated to the NR and Carroll JT gravities have been obtained using the method of non-linear realisations and IHM. In particular, we have shown that this reproduces the Drinfeld-Sokolov reduction of $SL(2,\mathbb R)$ \cite{Drinfeld:1984qv,Polyakov:1989dm}, which has been previously used in \cite{Saad:2019lba} to define boundary conditions of JT gravity in analogy to the three-dimensional gravity case \cite{Coussaert:1995zp}.
The procedure has been generalised to the case of the $SL(2,\mathbb R)\times \mathbb R$ algebra and the conformal descriptions of
NH$_2^{\pm}$ and Extended Carroll (A)dS$_2$. This leads to the NR Schwarzian action \eqref{NRSchwarzian} and its complexification \eqref{ComplexNRSchwarzian}. We have also derived the corresponding conserved charges and the asymptotic symmetry of these theories, which is given by the twisted Warped Virasoro algebra \cite{Afshar:2019axx,Hofman:2014loa,Godet:2020xpk}. Using the 
explicit form of the boundary gauge fields, we have reconstructed the fields in the bulk, which define Newton-Cartan \cite{Cartan:1923zea,Trautman:1963,Havas:1964zza,Kuenzle:1972zw,Kuchar:1980tw}
 and Carrollian structures \cite{Henneaux:1979vn,Duval:2014uoa}.   

Further studies can be done, for example:\\
The relation between the Carroll geometry \eqref{Carrollstructure} and Carrollian structures previously constructed in \cite{Henneaux:1979vn} and \cite{Duval:2014uoa}. The analysis here presented could be useful to find the asymptotic symmetries of general Newton-Cartan and Carroll geometries.  

Given the relation between IHM and Drinfeld-Sokolov reduction, it would be useful to do a systematic classification of all possible IHM and to study all possible boundary actions. 

An important future direction of this work is also the formulation of the post-Newtonian as well as post-Carrollian corrections of two-dimensional gravity and their boundary theories following recent results that have been obtained in three and four-dimensional gravity \cite{VandenBleeken:2017rij,Hansen:2018ofj,Bergshoeff:2019ctr,Ozdemir:2019orp,Gomis:2019nih}. This is a work in progress.

Other possible future directions are the study of the relation of the NR Schwarzian with a novel regime of the SYK model, the quantisation of the NR JT action and its Carrollian counterpart, their relation to matrix models along the lines of \cite{Saad:2019lba}, and their supersymmetric extensions.

\subsection*{Acknowledgments}

We thank Glenn Barnich, Jaume Gomis, Marc Henneaux, Axel Kleinschmidt, Mauricio Valenzuela and Jorge Zanelli for enlightening comments and discussions. We specially acknowledge Luis Avil\'es for taking active part in the initial stages of this work. JG has been supported in part by MINECO FPA2016-76005-C2-1-P 
and PID2019-105614GB-C21 and 
from the State Agency for Research of the
Spanish Ministry of Science and Innovation through the Unit of Excellence
Maria de Maeztu 2020-203 award to the Institute of Cosmos Sciences
(CEX2019-000918-M). DH is partially funded by ANID grant  \#21160649. The Centro de Estudios Cient\'ificos (CECs) is funded by the Chilean Government through the Centers of Excellence Base Financing Program of ANID. This work was partially supported by FNRS-Belgium (conventions FRFC PDRT.1025.14 and  IISN 4.4503.15), as well as by funds from the Solvay Family.

\appendix
\section{Conventions for (A)dS$_2$}
\label{AppAdS}
The (A)dS$_2$ algebra and its invariant bilinear form are defined as
\be\label{adsalginvt}
\bal
&[M_{AB},M_{CD}]=g_{AD} M_{BC}-g_{AC} M_{BD}-g_{BD} M_{AC}+g_{BC} M_{AD}\,,\\[5pt]
&\left\langle M_{AB},M_{CD}\right\rangle =\mu\left( g_{AD} g_{BC}-g_{AC} g_{BD}\right) \,,
\eal
\ee
where $A=0,1,2$, and $g_{AB}$ is the (A)dS$_2$ metric
\be\label{metricAdSgen}
g_{AB}={\rm diag}(-,+,-\sigma)
\ee
Here $\sigma=+1$ defines the AdS$_2$ algebra $\mathfrak so(2,1)$, while $\sigma=-1$ corresponds to the dS$_2$ case $\mathfrak so(1,2)$. Usually one redefines the generators in the (A)dS$_2$ basis as
\be\label{JPfromM}
J_{ab}=M_{ab}\,,\hskip .8truecm P_a=\sqrt{|\Lambda|} M_{2a}\,,\hskip.8truecm a=0,1\,,
\ee
where we have introduced the cosmological constant $\Lambda$ 
\be\label{cconstant}
\Lambda=-\sigma |\Lambda|
\ee
which defines (A)dS radius $R=1/\sqrt{|\Lambda|}$. The definitions \eqref{adsalginvt} then take the form
\be
\bal
&[J_{ab},P_{c}]=\eta_{bc} P_{a}-\eta_{ac} P_{b}\,,\hskip.8truecm
&[P_{a},P_{b}]=- \Lambda J_{ab}\,,\\[5pt]
&\left\langle J_{ab},J_{cd}\right\rangle = \mu\left(\eta_{ad} \eta_{bc}-\eta_{ac} \eta_{bd}\right)\,,\hskip.8truecm
&\left\langle P_{a},P_{b}\right\rangle=-\mu \Lambda\eta_{ab}
\eal
\ee
where $\eta_{ab}={\rm diag}(-,+)$ is the two-dimensional Minkowski metric. We can further simplify these relations by defining the dual generator
\be
J=-\frac{1}{2}\epsilon^{ab}J_{ab} 
\quad\Longrightarrow\quad J_{ab}=\epsilon_{ab} J\,, \hskip.5truecm  (\epsilon_{01}=-\epsilon_{10}=1)\,,
\ee
which leads to
\be\label{adsandpairing}
\bal
&[J,P_{a}]=\epsilon_a^{\;\;b} P_{b} \hskip.8truecm
&[P_{a},P_{b}]= - \Lambda \epsilon_{ab} J \,,\\[5pt]
&\left\langle J,J \right\rangle = \mu\,,\hskip.7truecm
&\left\langle P_{a},P_{b}\right\rangle=- \mu\Lambda\eta_{ab}
\eal
\ee

Both AdS$_2$ and dS$_2$ are isomorphic to the \asl algebra
\be\label{sl2ralg}
[\tilde{\mathcal D} , \tilde{\mathcal H}]= \tilde{\mathcal H}\,,\hskip.7truecm
[\tilde{\mathcal D}, \tilde{\mathcal K}]=- \tilde{\mathcal K}\,,\hskip.7truecm
[\tilde{\mathcal H}, \tilde{\mathcal K}]=2 \tilde{\mathcal D}
\ee

Setting $\sigma=-1$ in \eqref{cconstant}, we find that in the AdS$_2$ case we can define
\be\label{confbasisAdS}
\tilde{\mathcal D}=\ell P_1 \,,\hskip.8truecm
\tilde{\mathcal H}=\ell P_0+J \,,\hskip.8truecm
\tilde{\mathcal K}=\ell P_0-J \,
\ee
which leads to the \asl commutation relations \eqref{sl2ralg}. In the dS$_2$ case ($\sigma=1$), the conformal that leads to \asl reads
\be\label{confbasisdS}
\tilde{\mathcal D}=\ell P_0 \,,\hskip.8truecm
\tilde{\mathcal H}=\ell P_1 +J \,,\hskip.8truecm
\tilde{\mathcal K}=\ell P_1-J \,
\ee
In both cases the \asl pairing follows from \eqref{adsandpairing} and is given by
\be
\left\langle \tilde{\mathcal D} , \tilde{\mathcal D} \right\rangle=\mu\,,\hskip.7truecm
\left\langle \tilde{\mathcal H} , \tilde{\mathcal K} \right\rangle=-2\mu\,.
\ee
One can consider the following matrix representation
\be\label{sl2App}
\tilde{\mathcal D} =
\frac{1}{2}\begin{pmatrix}
1 & \phantom{-} 0 \\
0 & -1 
\end{pmatrix}
\,,\hskip.7truecm
\tilde{\mathcal K} =
\begin{pmatrix}
0 & -1 \\
0 & 0 
\end{pmatrix}
\,,\hskip.7truecm
\tilde{\mathcal H} =
\begin{pmatrix}
 0 & 0 \\
1 & 0 
\end{pmatrix}
\ee
for which the bilinear form can be expressed simply as $\left\langle\; \cdot  , \cdot \;\right\rangle = 2\mu \,{\rm Tr}(\;\;)$.

\section{Building blocks for the NR second order formulation}\label{solvinggammas}\label{AppendixSeconOrder}
In this appendix, we compute the NR second-order of JT gravity. The starting point is to consider a $\varepsilon$-expansion for the metric $g_{\mu \nu}$ \cite{Dautcourt:1997hb,Weinbergbook, dautcourt1964newtonske,DePietri:1994je,Hansen:2020pqs,VandenBleeken:2017rij,Ehlers:2019aco,Kuenzle:1972zw} and then from the metric compatibility, we find the Levi-Civita connections at the order in $\varepsilon$ which we are interested.  
\subsection{NR affine connections}\label{NRGammasapp}
Let us start with the expansion \eqref{transgmunu} of the metric and its inverse
\be \label{gmunuexpansion}
g_{\mu \nu} =  \frac{1}{\varepsilon^2} \,  {\overset{(2)} g}_{\mu \nu} +  {\overset{(0)} g}_{\mu \nu}, \hskip1.9truecm
g^{\mu \nu} =  {\overset{(0)} g}{}^{\mu \nu}+ \varepsilon^2 \, {\overset{(-2)} g}{}^{\mu \nu}\,.
\ee
From the relation $g_{\nu \rho} \, g^{\rho \mu} = \delta_\nu^\mu$, we get order-by-order the identities
\be \label{conditionsgammas}
 \overset{(2)}{g}_{\mu \rho} \overset{(0)}{g}{}^{\rho \nu} =  0\,, \qquad  \overset{(0)}{g}_{\nu \rho} \overset{(0)}{g}{}^{\rho \mu} + \overset{(2)}{g}_{\nu \rho} \overset{(-2)}{g}{}^{\rho \mu} = \delta_\nu^\mu\,, \qquad \overset{(0)}{g}_{\mu \rho} \overset{(-2)}{g}{}^{\rho \nu}=  0 \,.
\ee
Now we demand the metric compatibility condition for the Lorentzian metric, namely
\be\label{compaapp}
\nabla_\rho \, g_{\mu \nu} \, = \,  0\,.
\ee
The condition \eqref{compaapp} implies the following set of equations
\bseq\label{gamma-2gmunu0}
\ba 
\overset{(0)}{\nabla}_\rho    \overset{(0)}{g}_{\mu \nu}  -2\overset{(-2)}{\Gamma}{}^\lambda_{\rho (\nu} \overset{(2)}{g}_{\mu) \lambda} & = & 0\,, \\
 \overset{(0)}{\nabla}_\rho     \overset{(2)}{g}_{\mu \nu} -2\overset{(2)}{\Gamma}{}^\lambda_{\rho (\nu} \overset{(0)}{g}_{\mu) \lambda} & = & 0 \,, \\
 2\overset{(2)}{\Gamma}{}^\lambda_{\rho (\nu} \overset{(2)}{g}_{\mu) \lambda}  & = & 0 \,, \\
2 \overset{(-2)}{\Gamma}{}^{\lambda}_{\rho (\nu}  \overset{(0)}{g}_{\mu) \lambda} & = & 0\,.
\ea 
\eseq
In order to solve the affine connections, we proceed in the standard way by permuting indexes in \eqref{gamma-2gmunu0}, we find
\ba \label{eq1}
 &&\frac{1}{2} \left( \partial_\mu\, \overset{(2)}{g}_{\nu \rho}+ \partial_\nu\, \overset{(2)}{g}_{\mu \rho}- \partial_\rho\, \overset{(2)}{g}_{\mu \nu} \right)=\overset{(2)}{g}_{\rho \lambda} \overset{(0)}{\Gamma}{}^\lambda_{\mu \nu} +\overset{(0)}{g}_{\rho \lambda} \overset{(2)}{\Gamma}{}^\lambda_{\mu \nu}  \,,\\
  &&\frac{1}{2} \left( \partial_\mu\, \overset{(0)}{g}_{\nu \rho}+ \partial_\nu\, \overset{(0)}{g}_{\mu \rho}- \partial_\rho\, \overset{(0)}{g}_{\mu \nu} \right)=\overset{(0)}{g}_{\rho \lambda} \overset{(0)}{\Gamma}{}^\lambda_{\mu \nu} +\overset{(2)}{g}_{\rho \lambda} \overset{(-2)}{\Gamma}{}^\lambda_{\mu \nu} \,, \\
&& \overset{(2)}{\Gamma}{}^\lambda_{\mu \nu}\overset{(2)}{g}_{\rho \lambda} =  0\,,\\
&&   \overset{(-2)}{\Gamma}{}^\lambda_{\mu \nu}\overset{(0)}{g}_{\rho \lambda}  = 0\,.\label{gammaquasi-2}
\ea
Using the relations \eqref{conditionsgammas} in the last equations, we get the following expressions for the terms in the expansion Levi-Civita affine connection
\bseq\label{gammasappendix} 
\ba 
  \n \overset{(0)}{\Gamma}{}^\alpha_{\mu \nu}  & =&    \frac{1}{2}\overset{(0)}{g}{}^{\rho \alpha}  \left( \partial_\mu\, \overset{(0)}{g}_{\nu \rho}+ \partial_\nu\, \overset{(0)}{g}_{\mu \rho}- \partial_\rho\, \overset{(0)}{g}_{\mu \nu} \right) \\
  &&+\frac{1}{2}\overset{(-2)}{g}{}^{\rho \alpha}  \left( \partial_\mu\, \overset{(2)}{g}_{\nu \rho}+ \partial_\nu\, \overset{(2)}{g}_{\mu \rho}- \partial_\rho\, \overset{(2)}{g}_{\mu \nu} \right) \,, \\
\overset{(-2)}{\Gamma}{}^\alpha_{\mu \nu} & = &  \frac{1}{2}\overset{(-2)}{g}{}^{\rho \alpha}  \left( \partial_\mu\, \overset{(0)}{g}_{\nu \rho}+ \partial_\nu\, \overset{(0)}{g}_{\mu \rho}- \partial_\rho\, \overset{(0)}{g}_{\mu \nu} \right)  \,, \\
\overset{(2)}{\Gamma}{}^\alpha_{\mu \nu} & = &  \frac{1}{2}\overset{(0)}{g}{}^{\rho \alpha}  \left( \partial_\mu\, \overset{(2)}{g}_{\nu \rho}+ \partial_\nu\, \overset{(2)}{g}_{\mu \rho}- \partial_\rho\, \overset{(2)}{g}_{\mu \nu} \right) \,.
\ea
\eseq

\subsection{Ricci scalar} \label{DivergentRicciscalar}

Now we compute the different orders of the Ricci scalar expansion appearing in \eqref{Ricciexpansion}. In contrast to higher dimensions, we show that in two space-time dimensions the divergent terms vanish identically simply using the completeness relations of the fields of Newton-Cartan geometry.  

Starting from the standard relativistic formula of the Ricci scalar
\be
\mathcal{R}  =  g^{\mu \nu} \, (\partial_\lambda \, \Gamma^{\lambda}_{\mu \nu}  -\partial_\nu \, \Gamma^{\lambda}_{\lambda \mu}  + \Gamma^{\lambda}_{\lambda \alpha}\, \Gamma^{\alpha}_{\mu \nu} - \Gamma^{\lambda}_{\nu \alpha} \,\Gamma^{\alpha}_{\lambda \mu}) \,,
\ee
and substituting the expansions $\eqref{gmunuexpansion}$ and \eqref{gammasappendix}, we find the following expansion for the Ricci scalar
\be\label{Ricciexpansionapp}
\mathcal{R} = \frac{1}{\varepsilon^4} \,   \overset{(4)}{\mathcal R}  +  \frac{1}{\varepsilon^2} \,   \overset{(2)}{\mathcal R} +  \overset{(0)}{\mathcal R} + \varepsilon^2 \, \overset{(-2)}{\mathcal R}  + \mathcal{O} \left( \varepsilon^{4} \right)\,,
\ee
where
\ba
   \overset{(4)}{\mathcal R} & = &  2\, \overset{(0)}{g}{}^{\mu \nu} \, \overset{(2)}{\Gamma}{}^{\lambda}_{\alpha \lambda} \overset{(2)}{\Gamma}{}^{\alpha}_{\nu  \mu } \,,\\
   \overset{(2)}{\mathcal R} & = &   2 \overset{(-2)}{g}{}^{\mu \nu} \, \left( \overset{(2)}{\Gamma}{}^{\lambda}_{\alpha \lambda} \overset{(2)}{\Gamma}{}^{\alpha}_{\nu   \mu }    \right) +2 \overset{(0)}{g}{}^{\mu \nu}\, \left( \partial_{\lambda} {\overset{(2)}{\Gamma}{}^{\lambda}}_{\nu   \mu}  +\overset{(0)}{\Gamma}{}^{\lambda}_{\alpha \lambda } \overset{(2)}{\Gamma}{}^{\alpha}_{\nu   \mu} +\overset{(2)}{\Gamma}{}^{\lambda}_{\alpha  \lambda} \overset{(0)}{\Gamma}{}^{\alpha}_{\nu  \mu } 	  \right)\,, \\
    \n \overset{(0)}{\mathcal R} & = &  2\,  \overset{(-2)}{g}{}^{\mu \nu} \,  \left( \partial_{[\lambda}   \overset{(2)}{\Gamma}{}^{\lambda}_{\nu] \mu}  +  \overset{(0)}{\Gamma}{}^{\lambda}_{\alpha [\lambda}   \overset{(2)}{\Gamma}{}^{\alpha}_{\nu]  \mu}+ \overset{(2)}{\Gamma}{}^{\lambda}_{\alpha [\lambda}   {\overset{(0)}{\Gamma}^{\alpha}}_{\nu]  \mu}   \right) \\
 &&+2\,  \overset{(0)}{g}{}^{\mu \nu} \,  \left( \partial_{[\lambda} \overset{(0)}{\Gamma}{}^{\lambda}_{\nu] \mu}  +\overset{(0)}{\Gamma}{}^{\lambda}_{\alpha [\lambda} \overset{(0)}{\Gamma}{}^{\alpha}_{\nu] \mu} +\overset{(-2)}{\Gamma}{}^{\lambda}_{\alpha [\lambda} \overset{(2)}{\Gamma}{}^{\alpha}_{\nu] \mu } 	  +\overset{(2)}{\Gamma}{}^{\lambda}_{\alpha [\lambda} \overset{(-2)}{\Gamma}{}^{\alpha}_{\nu] \mu }  \right)\,, \\
 \n  \overset{(-2)}{\mathcal R} & = &     2\,  \overset{(-2)}{g}{}^{\mu \nu} \,   \, \left( \partial_{[\lambda}   \overset{(0)}{\Gamma}{}^{\lambda}_{\nu] \mu}  +  \overset{(0)}{\Gamma}{}^{\lambda}_{\alpha [\lambda}   \overset{(0)}{\Gamma}{}^{\alpha}_{\nu]  \mu} +\overset{(-2)}{\Gamma}{}^{\lambda}_{\alpha [\lambda}   \overset{(2)}{\Gamma}{}^{\alpha}_{\nu]  \mu} +\overset{(2)}{\Gamma}{}^{\lambda}_{\alpha [\lambda}   \overset{(-2)}{\Gamma}{}^{\alpha}_{\nu]  \mu}    \right) \\
&&+2\,  \overset{(0)}{g}{}^{\mu \nu} \,  \left( \partial_{[\lambda} \overset{(-2)}{\Gamma}{}^{\lambda}_{\nu] \mu}  +\overset{(0)}{\Gamma}{}^{\lambda}_{\alpha [\lambda} \overset{(-2)}{\Gamma}{}^{\alpha}_{\nu] \mu} +\overset{(-2)}{\Gamma}{}^{\lambda}_{\alpha [\lambda} \overset{(0)}{\Gamma}{}^{\alpha}_{\nu] \mu }  \right)\,.
\ea
Using the result $\overset{(2)}{\Gamma}{}^{\lambda}_{  \mu \nu}=0$ found in \eqref{gammasfromveilbeingpostulate}, the divergent terms vanish identically. The first term in the expansion in thus the finite term
\be\label{finiteRicci}
 \overset{(0)}{\mathcal R}  =  \overset{(0)}{g}{}^{\mu \nu} \left( \partial_{\lambda} \overset{(0)}{\Gamma}{}^{\lambda}_{\nu \mu}  -\partial_{\nu} \overset{(0)}{\Gamma}{}^{\lambda}_{\lambda \mu}  +\overset{(0)}{\Gamma}{}^{\lambda}_{\alpha \lambda} \overset{(0)}{\Gamma}{}^{\alpha}_{\nu \mu}-\overset{(0)}{\Gamma}{}^{\lambda}_{\alpha \nu} \overset{(0)}{\Gamma}{}^{\alpha}_{\lambda \mu} \right)\,.
\ee
Note that using \eqref{gammasfromveilbeingpostulate}, the terms with derivatives in \eqref{finiteRicci} take the form
\ba\label{derivativesfiniteRicci}
\n \partial_\lambda  \overset{(0)}{\Gamma}{}^{\lambda}_{\mu \nu} -\partial_\nu  \overset{(0)}{\Gamma}{}^{\lambda}_{\mu \lambda} & =& \partial_\lambda  \tau^\lambda \partial_\mu \tau_\nu  + \partial_\lambda  e^\lambda \left( \partial_\mu e_\nu + \omega_\mu \tau_\nu \right)  + e^\lambda \left( \partial_\lambda \omega_\mu \tau_\nu + \omega_\mu \partial_\lambda \tau_\nu\right)\\
&&- \partial_\nu  \tau^\lambda \partial_\lambda \tau_\mu  - \partial_\nu  e^\lambda \left( \partial_\lambda e_\mu + \omega_\lambda \tau_\mu \right)  + e^\lambda \left( \partial_\nu \omega_\lambda \tau_\mu + \omega_\lambda \partial_\nu \tau_\mu \right)\,.
\ea
Now, from the vielbein postulate \eqref{relativisticpostulate} together with the expressions \eqref{veilbein2ndorder} and \eqref{omegaexp} we get the following set of equations
\begin{multicols}{2}
\begin{subequations}
\setlength{\abovedisplayskip}{-13pt}
\allowdisplaybreaks
\begin{align}
\partial_\mu \, \tau_\nu - \overset{(0)}{\Gamma}{}^{\lambda}_{\mu \nu}\,  \tau_\lambda & =  0\,, \label{postulate1} \\[.2truecm]
\partial_\mu \, e_\nu 	+ \omega_\mu \, \tau_\nu - \overset{(0)}{\Gamma}{}^{\lambda}_{\mu \nu}\, e_\lambda &  = 0\,, \label{postulate2}\\[.1truecm]
\partial_\mu \, \tau^\nu - \omega_\mu \, e^\nu +\overset{(0)}{\Gamma}{}^{\nu}_{\mu \lambda}\,  \tau^\lambda & =   0\,, \label{postulate3}\\[.2truecm]
 \partial_\mu \, e^\nu + \overset{(0)}{\Gamma}{}^{\nu}_{\mu \lambda}\,  e^\lambda & =  0 \,.\label{postulate4}
\end{align}
\end{subequations}
\end{multicols}\noindent
Using \eqref{postulate2} and \eqref{postulate3}, we can rewrite some terms in \eqref{derivativesfiniteRicci} as follows
\ba
 \partial_\lambda  \tau^\lambda \, \partial_\mu \tau_\nu  + \partial_\lambda  e^\lambda \left( \partial_\mu e_\nu + \omega_\mu \tau_\nu \right) & = & \omega_\lambda \, e^\lambda \,  \overset{(0)}{\Gamma}{}^{\sigma}_{\mu \nu} \, \tau_\sigma -  \overset{(0)}{\Gamma}{}^{\lambda}_{ \lambda \rho}  \overset{(0)}{\Gamma}{}^{\rho}_{\mu \nu}\,, \label{intereq1}\\
 \partial_\nu  \tau^\lambda \, \partial_\lambda \tau_\mu  + \partial_\nu  e^\lambda \left( \partial_\lambda e_\mu + \omega_\lambda \tau_\mu \right) & = & \omega_\nu \, e^\lambda \,  \overset{(0)}{\Gamma}{}^{\sigma}_{\lambda \mu} \, \tau_\sigma -  \overset{(0)}{\Gamma}{}^{\lambda}_{ \nu \rho}  \overset{(0)}{\Gamma}{}^{\rho}_{\lambda \mu}\label{intereq2}\,.
\ea
This last fact is quite remarkable because helps us to cancel out the quadratic terms of the affine connection $\overset{(0)}{\Gamma}{}^{\lambda}_{\mu \nu}$ appearing in \eqref{finiteRicci}. Then, plugging \eqref{intereq1} and \eqref{intereq2} into \eqref{finiteRicci}, we find
\be
 \overset{(0)}{\mathcal R} = 0 \,.
\ee
Finally, we look at the last relevant term in the Ricci scalar expansion \eqref{Ricciexpansionapp}
\be\label{ricci-2}
\bal
\mathcal {\overset{\rm(-2)}R}&=
  2{\overset{(-2)}g}{}^{\mu\nu}\left( \partial_{[\lambda}  {\overset{(0)}\Gamma}{}^{\lambda}_{\nu] \mu}  +  {\overset{(0)}\Gamma}{}^{\lambda}_{\rho [\lambda} \, {\overset{(0)}\Gamma}{}^{\rho}_{\nu]  \mu}  \right) \\[4pt]
& +2{\overset{(0)}g}{}^{\mu\nu} \left( \partial_{[\lambda}{\overset{(-2)}\Gamma}{}^{\lambda}_{\nu] \mu}  +{\overset{(0)}\Gamma}{}^{\lambda}_{\rho [\lambda}\,{\overset{(-2)}\Gamma}{}^{\rho}_{\nu] \mu} +{\overset{(-2)}\Gamma}{}^{\lambda}_{\rho [\lambda}\, {\overset{(0)}\Gamma}{}^{\rho}_{\nu] \mu }  \right)
\,.
\eal
\ee
The terms involving ${\overset{(0)}\Gamma}{}^{\lambda}_{\mu \nu}$ are computed in the same way as done before. This yields
\be
2{\overset{(-2)}g}{}^{\mu\nu}\left( \partial_{[\lambda}  {\overset{(0)}\Gamma}{}^{\lambda}_{\nu] \mu}  +  {\overset{(0)}\Gamma}{}^{\lambda}_{\rho [\lambda} \, {\overset{(0)}\Gamma}{}^{\rho}_{\nu]  \mu}  \right)  =   2 \, \tau^{\mu} \, e^{\nu} \, \partial_{[\mu} \, \omega_{\nu]}\,.
\ee
 We now need to compute the terms proportional to ${\overset{(-2)}\Gamma}{}^{\lambda}_{\mu \nu}$. In order to do so, let us consider the following formulae obtained from the using \eqref{inspiredcontraction} and \eqref{omegaexp} in the relativistic inverse vielbein postulate 
 \bseq\label{identitiesRmenos2}
\ba
 \partial_\mu \tau^\nu - \omega_\mu e^\nu +  {\overset{(0)}\Gamma}{}^{\nu}_{\mu \rho} \tau^\rho & = & 0\,,\label{post1} \\
 \partial_\mu e^\nu +  {\overset{(0)}\Gamma}{}^{\nu}_{\mu \rho} e^\rho & =  & 0 \,, \label{post2}\\
\omega_\mu \tau^\nu -   {\overset{(0)}\Gamma}{}^{\nu}_{\mu \rho} e^\rho &  = & 0 \,, \label{post3} \\
\partial_\mu m_\nu + \omega_\mu e_\nu - {\overset{(0)}\Gamma}{}^{\rho}_{\mu \nu} m_\rho- {\overset{(-2)}\Gamma}{}^{\rho}_{\mu \nu} \tau_\rho & =  & 0 \,,\label{post4} \\
\omega_\mu m_\nu -  {\overset{(-2)}\Gamma}{}^{\rho}_{\mu \nu} e_\rho & = & 0\,.\label{post5}
\ea
\eseq
We start computing the term,
\ba\label{firstderiva}
\n &&\partial_\lambda \,  {\overset{(-2)}\Gamma}{}^{\lambda}_{\mu \nu}  =  \partial_\lambda \tau^\lambda \left(   \partial_\mu m_\nu + \omega_\mu e_\nu -  {\overset{(0)}\Gamma}{}^{\sigma}_{\mu \nu} m_\sigma     \right)  \\
\n &&+ \tau^\lambda  \left( \partial_\lambda \partial_\mu m_\nu  + \partial_\lambda \omega_\mu e_\nu + \omega_\mu \partial_\lambda e_\nu - \partial_\lambda  {\overset{(0)}\Gamma}{}^{\sigma}_{\mu \nu} m_\sigma - {\overset{(0)}\Gamma}{}^{\sigma}_{\mu \nu} \partial_\lambda m_\sigma \right) \\
&& +\partial_\lambda e^\lambda \omega_\mu m_\nu + e^\lambda \left( \partial_\lambda \omega_\mu m_\nu + \omega_\mu \partial_\lambda m_\nu \right) \,.
\ea
Using \eqref{post1} and \eqref{post4}, we can rewrite the first term on the right hand side of \eqref{firstderiva} as
\ba\label{term1}
\n \partial_\lambda \tau^\lambda \left(   \partial_\mu m_\nu + \omega_\mu e_\nu -  {\overset{(0)}\Gamma}{}^{\sigma}_{\mu \nu} m_\sigma     \right) & = & \left( \omega_\lambda e^\lambda -  {\overset{(0)}\Gamma}{}^{\lambda}_{\lambda \rho} \tau^\rho \right)  {\overset{(-2)}\Gamma}{}^{\sigma}_{\mu \nu}\tau_\sigma \,, \\
 & = & \omega_\lambda e^\lambda {\overset{(-2)}\Gamma}{}^{\sigma}_{\mu \nu}\tau_\sigma - {\overset{(0)}\Gamma}{}^{\lambda}_{\lambda \rho}  {\overset{(-2)}\Gamma}{}^{\rho}_{\mu \nu} + {\overset{(0)}\Gamma}{}^{\lambda}_{\lambda \rho}  {\overset{(-2)}\Gamma}{}^{\sigma}_{\mu \nu} e^\rho e_\sigma \,,
\ea
where we have used the completeness relation \eqref{S1invertconditions}.
Using \eqref{post4}, we can rewrite the last term in the second line of \eqref{firstderiva}, namely
\ba\label{term2}
 \n \tau^\lambda  {\overset{(0)}\Gamma}{}^{\sigma}_{\mu \nu}  \partial_\lambda m_\sigma &  =& \tau^\lambda  {\overset{(0)}\Gamma}{}^{\sigma}_{\mu \nu}  \left(   {\overset{(0)}\Gamma}{}^{\rho}_{\lambda \sigma} m_\rho +  {\overset{(-2)}\Gamma}{}^{\rho}_{\lambda \sigma} \tau_\rho - \omega_\lambda e_\sigma \right) \\
 & = & \tau^\lambda  {\overset{(0)}\Gamma}{}^{\sigma}_{\mu \nu}  \,   {\overset{(0)}\Gamma}{}^{\rho}_{\lambda \sigma} m_\rho  +  {\overset{(0)}\Gamma}{}^{\rho}_{\mu \nu} {\overset{(-2)}\Gamma}{}^{\lambda}_{\lambda \rho} -{\overset{(0)}\Gamma}{}^{\sigma}_{\mu \nu} {\overset{(-2)}\Gamma}{}^{\rho}_{\lambda \sigma} e^\lambda e_\rho-  {\overset{(0)}\Gamma}{}^{\sigma}_{\mu \nu} \tau^\lambda \omega_\lambda e_\sigma \,.
\ea
Plugging \eqref{term1} and \eqref{term2} into \eqref{firstderiva}, we get
\ba
\n  \partial_\lambda \,  {\overset{(-2)}\Gamma}{}^{\lambda}_{\mu \nu}   & = & \omega_\lambda e^\lambda {\overset{(-2)}\Gamma}{}^{\sigma}_{\mu \nu}\tau_\sigma - {\overset{(0)}\Gamma}{}^{\lambda}_{\lambda \rho}  {\overset{(-2)}\Gamma}{}^{\rho}_{\mu \nu} + {\overset{(0)}\Gamma}{}^{\lambda}_{\lambda \rho}  {\overset{(-2)}\Gamma}{}^{\sigma}_{\mu \nu} e^\rho e_\sigma \\
\n && + \tau^\lambda  \left( \partial_\lambda \partial_\mu m_\nu  + \partial_\lambda \omega_\mu e_\nu + \omega_\mu \partial_\lambda e_\nu - \partial_\lambda  {\overset{(0)}\Gamma}{}^{\sigma}_{\mu \nu} m_\sigma \right) \\
 \n &&- \tau^\lambda  {\overset{(0)}\Gamma}{}^{\sigma}_{\mu \nu}  \,   {\overset{(0)}\Gamma}{}^{\rho}_{\lambda \sigma} m_\rho  -  {\overset{(0)}\Gamma}{}^{\rho}_{\mu \nu} {\overset{(-2)}\Gamma}{}^{\lambda}_{\lambda \rho} +{\overset{(0)}\Gamma}{}^{\sigma}_{\mu \nu} {\overset{(-2)}\Gamma}{}^{\rho}_{\lambda \sigma} e^\lambda e_\rho+   {\overset{(0)}\Gamma}{}^{\sigma}_{\mu \nu} \tau^\lambda \omega_\lambda e_\sigma \\
&& +\partial_\lambda e^\lambda \omega_\mu m_\nu + e^\lambda \left( \partial_\lambda \omega_\mu m_\nu + \omega_\mu \partial_\lambda m_\nu \right) \,.
\ea
Replacing these results in \eqref{ricci-2} a little algebra leads to
 \ba
\n && \overset{\rm(-2)}{\mathcal{R}} =  4 \, \tau^\mu e^\nu \partial_{[\mu} \omega_{\nu ]} \\
 && +\, h^{\mu \nu} \left(  e^\rho e_\sigma \left[  {\overset{(0)}\Gamma}{}^{\lambda}_{\lambda \rho} {\overset{(0)}\Gamma}{}^{\sigma}_{\mu \nu} - {\overset{(0)}\Gamma}{}^{\lambda}_{\nu \rho} {\overset{(0)}\Gamma}{}^{\sigma}_{\lambda \mu}         \right]           +\partial_\lambda e^\lambda \omega_\mu m_\nu - \partial_\nu e^\lambda \omega_\lambda m_\mu  \right) \,.
\ea
Finally, using the identities \eqref{identitiesRmenos2}, we find
\be
 \overset{\rm(-2)}{\mathcal{R}} =  4 \, \tau^\mu e^\nu \partial_{[\mu} \omega_{\nu ]}\,.
\ee

\end{document}